# Can Protostellar Outflows Set Stellar Masses?

*short title:* Protostellar Outflows

Philip C. Myers
Center for Astrophysics | Harvard and Smithsonian (CfA),
60 Garden Street
Cambridge, MA 02138, USA
pmyers@cfa.harvard.edu

Michael M. Dunham
Department of Physics, State University of New York at Fredonia
280 Central Avenue
Fredonia, NY 14063, USA

Ian W. Stephens
Department of Earth, Environment, and Physics
Worcester State University
486 Chandler St.
Worcester, MA 01602, USA

**Abstract**

The opening angles of some protostellar outflows appear too narrow to match the expected core-star mass efficiency $SFE = 0.3 - 0.5$, if outflow cavity volume traces outflow mass, with a conical shape and a maximum opening angle near 90 degrees. However, outflow cavities with paraboloidal shape and wider angles are more consistent with observed estimates of the *SFE*. This paper presents a model of infall and outflow evolution based on these properties. The initial state is a truncated singular isothermal sphere which has mass $\approx 1\ M_\odot$, free fall time $\approx 80$ kyr, and small fractions of magnetic, rotational, and turbulent energy. The core collapses pressure-free as its protostar and disk launch a paraboloidal wide-angle wind. The cavity walls expand radially and entrain envelope gas into the outflow. The model matches *SFE* values when the outflow mass increases faster than the protostar mass by a factor 1-2, yielding protostar masses typical of the IMF. It matches observed outflow angles if the outflow mass increases at nearly the same rate as the cavity volume. The predicted outflow angles are then typically ~50 deg as they increase rapidly through the stage 0 duration of ~40 kyr. They increase more slowly up to ~110 deg during their stage I duration of ~70 kyr. With these outflow rates and shapes, model predictions appear consistent with observational estimates of typical stellar masses, *SFEs*, stage durations, and outflow angles, with no need for external mechanisms of core dispersal.

# 1. Introduction

Many low-mass Class 0 protostars are born single in dense cores, according to recent high-resolution surveys (Tobin et al. 2022, Offner et al. 2021; see also Lada 2006). Nearly all such protostars drive molecular outflows of roughly bipolar shape, with each lobe approximated as a cone or paraboloid (Snell et al. 1980; Arce et al. 2006, hereafter A06; Bally 2016). The outflow may arise as a jet or a wide-angle wind entrains envelope gas (Pudritz & Norman 1986, Frank et al. 2014, Zhang et al. 2016, hereafter Z16, Zhang et al. 2019, hereafter Z19). Outflow cavity angles appear to widen with increasing protostar age, from the embedded class 0 phase to the disk-dominated class I phase (Arce & Sargent 2006, Offner et al. 2011, Velusamy et al. 2014, hereafter V14, Hsieh et al. 2017, Dunham et al. 2023, hereafter D23).

Mass loss due to such outflows has long been considered a way to clear the protostellar envelope and to set the low efficiencies of star formation in dense cores and in molecular clouds (Nakano et al. 1995; Matzner & McKee 2000, hereafter MM00, Hansen et al. 2012, Cunningham et al. 2018, Rohde et al. 2021). The efficiency in dense cores (*SFE* or $\epsilon$) is the ratio of the final mass of the protostar-disk system to the core mass at the start of collapse. Typical estimates of $\epsilon$ are 0.3 - 0.5 (Alves et al. 2007, Enoch et al. 2008, Könyves et al. 2015, Könyves et al. 2020) based on the shift between the initial mass function of stars (IMF) and core mass functions (CMFs) in nearby star-forming clouds. Cores considered likely to form stars are bound "prestellar" cores or "phase III" cores (Offner et al. (2022a).

Not all core population studies find $\epsilon$ = 0.3 - 0.5. A high-resolution study of Orion cores and stars finds mass functions whose power-law slopes are nearly equal to that of the IMF. Their turnover masses are also nearly equal, at ~ 0.2 $M_\odot$ (Takemura et al. 2023). This result would imply $\epsilon \approx 1$ in the standard *SFE* analysis. Such high efficiency is inconsistent with star-forming infall and outflow from a fixed-mass core. However it may be consistent with infall and outflow from cores having sufficiently high accretion rates, as discussed in Section 7.2.2.

Recent observations of outflow structure have raised questions about the core-clearing role of outflows from fixed-mass cores. The full-width opening angles of outflow cavities in CO lines and in the mid-infrared are typically less than 80 - 110 deg, as discussed in Section 4.3. These angles may be too small to match estimates of *SFE*, assuming that the outflow cavity volume traces the outflow mass, and that the cavity has conical shape (AS06, D23 (Habel et al. 2021, hereafter H21; D23). If so, the mass loss due to outflows may be insufficient to terminate accretion and to set typical values of protostar mass.

If outflows cannot clear enough core gas, other mechanisms of core gas dispersal may be needed, including feedback from winds, radiation, and ionization from nearby young stars. These external mechanisms of dispersal are often considered in regions of more massive star formation (Hosokawa et al. 2011, Kuiper et al. 2016, Tanaka et al. 2017).

On the other hand, departures from conical shape and/or departures from the equality of outflow mass fraction and volume fraction may improve the consistency between outflow angles and estimates of *SFE*.

Consistency with the *SFE* may improve if the outflow cavity base is significantly wider than the width of a cone, so that the outflow removes more envelope mass for a given opening angle. The volume of a cavity whose height $z$ increases with width $x$ as $z \propto x^n$ increases as the degree $n$ of the power law increases, where $n = 1$ for a cone and $n = 2$ for a paraboloid. Scattered-light outflow cavities observed at near infrared (NIR) wavelengths have been fit with such power-law models, indicating a mean degree $\bar{n} = 1.9$, with some degree values as great as $n = 6$ (H21). Models of wide-angle winds in magnetized cores predict paraboloidal outflow shapes (Li & Shu 1996), and these models have been used to interpret observations of CO outflows (Lee et al. 2000). ALMA observations of CO outflows in HH 46/47 are well-fit by models of expanding paraboloidal shells (Z16, Z19), and their associated time scales also indicate consistency with estimates of *SFE* (Z16).

Consistency with the *SFE* may also improve if the outflow cavity volume grows more slowly than the total outflow mass. Such a difference in growth rates could arise if the infalling envelope provides increasing resistance to cavity expansion. Also, the core magnetic field may limit the cavity growth across field lines, according to simulations of rotating, collapsing, magnetized cores (Machida & Matsumoto 2012, hereafter MM12; Machida & Hosokawa 2013, hereafter MH13). In these simulations the outflow removes significant core mass despite its relatively small width. The predicted cavity opening angle on the core scale increases to ~ 90 deg during the stage 0 phase. During the stage I phase the wind continues to entrain and expel envelope gas with constant cavity width on the core scale.

The outflows described by MM12 and MH13 differ from those of Z16 and Z19 in their gas entrainment and in the prevalence of their radial motions. In MM12 and MH13 the outflow entrains envelope gas close to the disk. This entrainment requires more nonradial envelope motion in order to join the outflow (MH13 Figures 5 and 6). In Z16 and Z19 the expanding cavity entrains envelope gas at all heights along the cavity wall, as the wall expands radially outward. Both types of entrainment may be consistent with observations (Z19), but radial motions are more compatible with the spherically symmetric initial model to be adopted here. Thus the assumed core geometry in Figure 1 is based on radial infall, and on radial expansion of paraboloidal outflow shells.

If outflow models match observed estimates of opening angles and the *SFE*, it is also important to determine whether such models are consistent with observed protostar and core masses, with the durations of their evolutionary stages, and with the evolution of their outflow angles. This paper addresses these points with a simple model of infall and outflow (hereafter IO model) in a low-mass dense core.

The initial state of the IO model resembles a singular isothermal sphere (SIS; Shu 1977) with relatively small fractions of magnetic, rotational, and turbulent energy. The core collapses radially inward to form a disk and protostar, which launch episodic wide-angle wind shells. The shells merge and push radially outward into the infalling envelope (Z16, Z19). In this model radial motions predominate except on the scales of the disk and the cavity-envelope interface, which are negligibly small compared to the core scale.

The structure of this paper is as follows. Section 2 derives expressions for the mass fractions of the protostar, envelope, and outflow, as functions of time and of the dispersal parameter $\alpha \equiv \tau_d/\tau_f$. Here $\alpha$ is the dispersal time scale $\tau_d$ normalized by the initial free-fall time $\tau_f$. Section 3 presents the time history of these component masses depending on $\alpha$. Section 4 describes observed outflow opening angles and presents cavity volume fractions for power-law cavity shapes. It predicts outflow angle evolution for a paraboloidal cavity expanding linearly with time as in Z19. Section 5 compares predicted masses, efficiencies, class durations, and outflow angles with observations. Section 6 summarizes the model. Section 7 discusses and interprets the results, and section 8 presents the conclusion.

## 2. Infall-Outflow Model

This infall-outflow model (hereafter "IO model") is an idealized description of a low-mass star-forming dense core, with typical initial mass 1 $M_\odot$, temperature 10 K, and radius $6.5 \times 10^3$ au. Its mass components include the infalling envelope, the protostellar star-disk system, and the outflowing gas. The initial envelope gas has the density structure of a singular isothermal sphere (SIS; Chandrasekhar 1939, Shu 1977, hereafter S77). It is truncated at radius $a$ by a low-density medium of constant pressure. Its magnetic and rotational energies are at most a few percent of its gravitational energy. They are small enough to neglect in the initial core-scale envelope structure, but sufficient to generate a hydromagnetic disk wind during protostar formation (MM12, MH13). The envelope velocity dispersion has equal contributions from thermal and microturbulent motions, approximating dense core observations in spectral lines of $NH_3$ (Chen et al. 2019). The relatively low mass of the protostellar jet and the disk are neglected, and the "protostar-disk system" is called the "protostar" for simplicity. The wide-angle outflow cavity associated with the disk wind is idealized as a hollow bipolar paraboloid of full opening angle $\phi_{oa}$.

Figure 1 is a sketch of the model components in the initial state and after half of the initial free-fall time. At this time, the outer radius has contracted by about 10%. The circular black lines indicate an infalling mass shell at half the initial radius, with inward radial arrows indicating a speed ~ 1 km s$^{-1}$. The parabolic shell lines indicate a bipolar outflow shell (Lee et al. 2000, Z19), with outward radial arrows indicating shell speed as given in Z19. For simplicity the central concentration and flattening of the density structure are not shown.

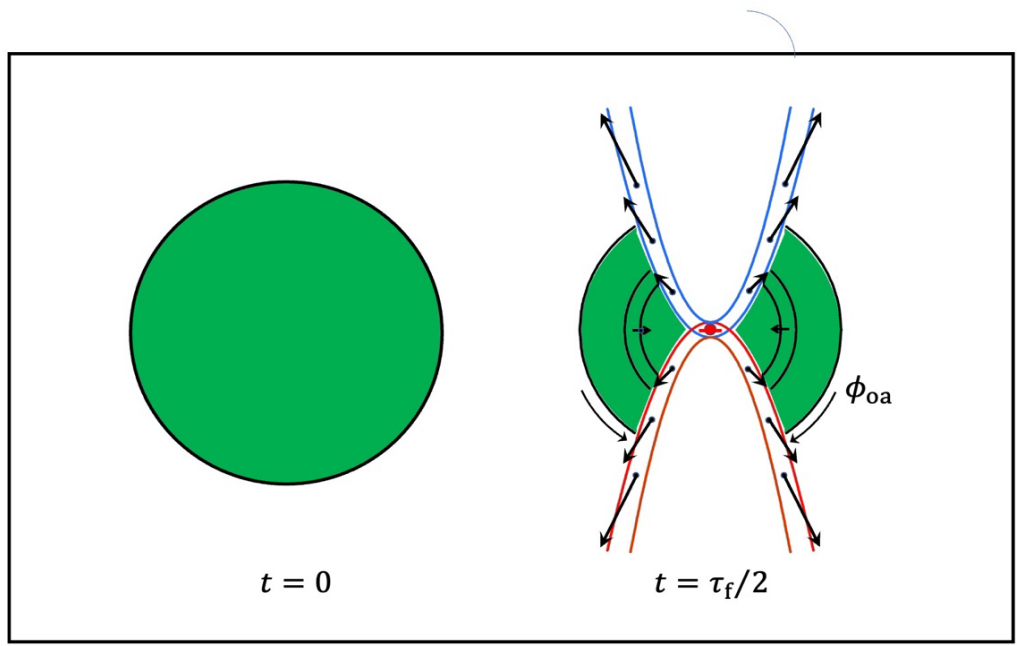

**Figure 1.** Sketch of IO model components in a plane containing the outflow axis and protostar at the center of a spherical initial core. For simplicity the central concentration of the core gas is not shown. The left panel shows the initial core. The right panel shows the core components after half of a free-fall time, at the end of the Stage 0 phase. These components include the central protostar and disk, an infalling spherical shell at half the initial radius (*black*), and paraboloidal outflow shells (*red* and *blue*) due to wide-angle winds as in Lee et al. (2000), Z16 and Z19. Each outflow shell spans the opening angle $\phi_{oa}$ evaluated at the core outer radius. *Inward arrows* show the radial inward motion of the infalling spherical shell. *Outward arrows* show the radial motion of the paraboloidal outflow shells.

The core evolution depends on the free fall time $\tau_f$ within its initial truncation radius, on the dispersal time scale $\tau_d$ for envelope mass loss to the outflow, and on the "infall time" $\tau_{in}$ for the outermost shell to fall to the origin. Here $\tau_{in} > \tau_f$ due to dense gas dispersal by the outflow. The "dispersal parameter" $\alpha \equiv \tau_d/\tau_f$ indicates the relative importance of infall and dispersal.

In the following, section 2.1 derives the protostar mass accretion rate. Section 2.2 derives the mass evolution of the protostar, the infalling envelope, and the outflow mass. Section 2.3 calculates $\tau_{in}$ by numerical solution of the equation of motion of an infalling shell. Section 2.4 quantifies the resulting relations between $\alpha$, $\tau_{in}$, and $\epsilon$. Section 2.5 uses these results to predict final mass fractions of the protostar, envelope, and outflow as functions of $\alpha$. Section 2.6 describes the relative rates of envelope mass flow to the outflow and to the protostar.

## 2.1. Protostellar Mass Accretion Rate

The protostellar mass accretion rate $dm_{ps}/dt$ is derived from the rate at which a shell of envelope gas falls from its initial radius to accrete onto the protostar. The envelope consists of all gas within the initial core boundary which is not in the outflow and not in the protostar.

The core free fall time is denoted as $\tau_f$ from the initial truncation radius $a$, and as $\tau_{fa'}$ from initial interior shells of radius $a'$, where $0 \leq a' < a$. The rate of envelope mass loss to the outflow is assumed to be $dm_{env}/dt = -m_{env}/\tau_d$, for consistency with tapered mass accretion rates derived from outflow observations (Bontemps et al. 1996, hereafter B96; Curtis et al. 2010), protostar surveys (Fischer et al. 2017), and simulations (Schmeja & Klessen 2004, Vorobyov 2010). Thus after an envelope shell falls inward for time $t = \tau_d$, it has lost a fraction $e^{-1}$ of its initial mass to the outflow.

The protostar mass accretion rate is a modified version of the rate for the pressure-free collapse of a SIS which is truncated at outer radius $a$, with enclosed mass $M(0)$. Pressure-free collapse can approximate isothermal collapse because during isothermal collapse, gravity dominates pressure by an increasing factor (Krumholz 2015).

In the pressure-free collapsing SIS, a falling shell which reaches the origin at time $t$ has the initial radius $a'$ for which $t$ equals the SIS free fall time, i.e. $t = \tau_{fa'} = \pi a'/(4\sigma)$ where $\sigma$ is the initial velocity dispersion. This shell has origin-crossing time interval $dt = \pi da'/(4\sigma)$ and mass $dm = 2\sigma^2 da'/G$ for shell thickness $da'$. Thus the protostar mass accretion rate is $dm_{ps}/dt = 8\sigma^3/(\pi G)$. This rate is constant in time and independent of initial radius.. It is greater by a factor of order unity than the mass accretion rate of a SIS collapsing with thermal pressure (S77). Since $8\sigma^3/(\pi G) = M(0)/\tau_f$, the protostar mass accretion rate is constant for $0 \leq t \leq \tau_f$,

$$\frac{dm_{ps}}{dt} = \frac{M(0)}{\tau_f} \quad . \tag{1}$$

When the pressure-free collapsing SIS has outflow mass loss, each infalling spherical shell has a high density zone and two low density zones which channel the bipolar outflowing gas. In the IO model of this system, each such nonuniform shell is approximated by a uniform spherical shell of the same radius, thickness, and mean density. For this shell, the infall time exceeds its initial free-fall time, because the mass loss of the shell interior reduces the inward gravitational acceleration of the shell. Infall times are calculated by solving the shell equation of motion for a wide range of values of $\alpha$ in Section 2.3.

For a given value of $\alpha$, the ratio of infall time to free fall time for each interior shell is assumed to be the same as the ratio for the outermost shell, due to the scale-free density structure of the SIS. The ratio of infall to free fall time is denoted as $\theta_{in} \equiv \tau_{in}/\tau_f$.

Outflow mass loss also causes the infalling shell mass to decrease with time $t$, by the same factor $\exp(t/\tau_d)$ as assumed above for the shell interior. Thus the protostar mass accretion rate for the collapsing, mass-losing SIS equals the rate for the collapsing, constant-mass SIS in equation (1), multiplied by the factor $\theta_{in}^{-1}\exp(-t/\tau_d)$, or

$$\frac{dm_{ps}}{dt} = \frac{M(0)}{\tau_f \theta_{in}} \exp(-t/\tau_d) \quad . \tag{2}$$

In terms of normalized quantities $\mu_{ps} \equiv m_{ps}/M(0)$, $\theta \equiv t/\tau_f$, and $\alpha \equiv \tau_d/\tau_f$, the mass accretion rate in equation (2) has the normalized form

$$\frac{d\mu_{ps}}{d\theta} = \frac{\exp(-\theta/\alpha)}{\theta_{in}} \quad , \tag{3}$$

for normalized times $0 \leq \theta \leq \theta_{in}$. This expression differs by a factor of $\alpha$ from that of B96 Section 4.3.1, because here $\tau_d$ and $\tau_{in}$ are allowed to differ, while B96 assumed that they are equal. A typical value of $\alpha$ is $\alpha \approx 0.5$ as shown in section 2.5.

## 2.2. Core Component Mass Fractions

This section gives expressions for the component masses of the protostellar dense core, normalized by the initial core mass $M(0)$, as a function of normalized time $\theta$ since the start of infall, in terms of the parameters $\alpha$ and $\theta_{in}$ defined above.

The normalized protostar mass at time $\theta$ is obtained by integration of equation (3) over time,

$$\mu_{ps} = \frac{\alpha}{\theta_{in}} [1 - \exp(-\theta/\alpha)] \quad . \tag{4}$$

The protostar accretion stops when the outermost shell has fallen to the origin, when $\theta = \theta_{in}$, after which time the protostar mass is assumed to remain constant. Then $\mu_{ps}$ has the final value $\mu_{ps,f}$, which is identical with the *SFE*, denoted in Section 1 as $\epsilon$:

$$\epsilon = \frac{\alpha}{\theta_{in}} [1 - \exp(-\theta_{in}/\alpha)] \quad . \tag{5}$$

The envelope mass at time $t$ is derived in terms of the initial radii which bound the gas which has neither accreted onto the protostar nor dispersed to the outflow,

$$M_{env} = \frac{2\sigma^2[a-a'(t)]}{G}\exp(-t/\tau_d) . \quad (6)$$

Here $a$ is the outermost radius of the initial core, and $a'(t)$ is the initial radius of the shell which crosses the origin at time $t$, $a'(t) = 4\sigma t/(\pi\theta_{in})$. The initial gas between $a$ and $a'(t)$ has mass $2\sigma^2[a-a'(t)]/G$, since $2\sigma^2 a/G$ is the initial SIS mass enclosed by radius $a$. At time $t$ this gas has lost a fraction $[1-\exp(-t/\tau_d)]$ of its initial mass to the outflow, leaving the surviving fraction $\exp(-t/\tau_d)$ in the infalling envelope. Applying the same normalizations as in Section 2.1 to equation (6) yields the envelope mass fraction,

$$\mu_{env} = \left(1 - \frac{\theta}{\theta_{in}}\right)\exp(-\theta/\alpha) \quad (7)$$

for $0 \leq \theta \leq \theta_{in}$. When $\theta \geq \theta_{in}$, $\mu_{env} = 0$ since all of the envelope gas has either dispersed through the outflow or accreted onto the protostar.

The mass fraction of gas in the outflow is obtained by conservation of the initial core mass in the form of protostar, envelope, and outflow, i.e.

$$\mu_{out} = 1 - \mu_{ps} - \mu_{env} \quad (8)$$

for $0 \leq \theta \leq \theta_{in}$. Note that $\mu_{out}$ includes both wind gas and entrained gas, and that $\mu_{out}$ is dominated by entrained gas (e.g. Watson et al. 2016). When $\theta \geq \theta_{in}$, $\mu_{env} = 0$. Then from equation (8) the final outflow mass fraction can be written in terms of the final protostar mass fraction, $\mu_{out,f} = 1 - \mu_{ps,f}$, or equivalently

$$\mu_{out,f} = 1 - \epsilon . \quad (9)$$

The enclosed mass fraction within the outermost mass shell, including protostar mass and envelope mass, is written as

$$\mu_{em} \equiv \mu_{ps} + \mu_{env} \quad (10)$$

when $0 \leq \theta \leq \theta_{in}$. This quantity $\mu_{em}$ is needed to obtain the infall time $\theta_{in}$ as a function of $\alpha$ in Section 2.3. As $\theta$ increases from 0 to $\theta_{in}$, $\mu_{em}$ decreases monotonically from 1 to $\epsilon$, due to the increase in $\mu_{ps}$ and the decrease in $\mu_{env}$.

### 2.3. Infall Time

The net loss of mass enclosed by an infalling shell continuously decreases the shell's inward gravitational acceleration, so its infall time $\tau_{in}$ exceeds its initial free fall time $\tau_f$. It is necessary to quantify the dependence of $\tau_{in}$ on $\alpha$, to predict the evolution of protostellar systems with dispersing envelopes. Here $\tau_{in}$ is calculated from a modification of the pressure-free collapse of a sphere of constant mass (Hunter 1962, hereafter H62; Spitzer 1978; Krumholz 2015).

The equation of motion (EOM) for a pressure-free collapsing sphere of constant mass $M$ is written

$$\frac{d^2r}{dt^2} = -\frac{GM}{r^2} \qquad (11)$$

where $r$ is the radius of a shell at time $t$ since the start of collapse, and $a$ is its initial radius at $t = 0$. The sphere mass $M$ is independent of time, and the initial mean density is $\bar{\rho}(a) = 3M/(4\pi a^3)$. The time for a shell to fall from radius $a$ to the center is the free-fall time, $\tau_f \equiv [3\pi/(32G\bar{\rho}(a))]^{1/2}$. This time has a simple analytic expression because the EOM can be solved analytically (H62).

When the mass of the collapsing sphere decreases with time due to protostellar outflow, the mass in equation (11) is written $M = M(0)\mu_{em}$. Here $M(0)$ is the initial mass and $\mu_{em}$ in equation (10) decreases with time from 1 to $\epsilon$ as noted above. The corresponding EOM does not have a known analytic solution, even when the dependence of $\mu_{em}$ on $t$ is independent of radius (Polyanin & Zaitsev 2018). Instead, the EOM

$$\frac{d^2r}{dt^2} = -\frac{GM(0)}{r^2}\mu_{em} \qquad (12)$$

is solved here numerically.

The enclosed mass fraction $\mu_{em}$ given in equation (10) depends on $\theta$, $\alpha$, and $\theta_{in}$. For each value of $\alpha$, the input value of $\theta_{in}$ in $\mu_{em}$ is adjusted until it matches the output value of $\theta$ in the numerical solution to equation (12), when $r = 0$. The resulting values of $\theta_{in}$ should meet the requirements (1) when $\alpha$ decreases, $\theta_{in}$ must increase, since a progressively faster outflow reduces the mass interior to a shell, which reduces the inward gravitational force on the shell; and (2) when $\alpha \to \infty$ and the outflow mass loss becomes negligible, the ratio $r(t)/a$ must exactly match the H62 solution.

To solve the EOM, equation (12) is written in the dimensionless form

$$\frac{d^2 y}{d\theta^2} = -\frac{\pi^2}{8y^2}\mu_{em} \quad . \tag{13}$$

Here $y \equiv r/a$ is the shell radius normalized by its initial value and $\theta \equiv t/\tau_f$ is the time since the start of collapse, normalized by the initial free-fall time. Equation (13) was solved using the fourth-order Runge-Kutta method, for $0.10 \leq \alpha \leq \infty$, with time step intervals $\Delta\theta$ from 0.01 to 0.05. The initial conditions are $y(0) = 1$ since $a$ is the initial value of $r$, and $dy(0)/d\theta = 0$, since the collapse starts from rest. The solutions were calculated with the Wolfram|Alpha Notebook Edition software package (https://www.wolfram.com/wolfram-alpha-notebook-edition). To our knowledge this solution has not been reported previously. A selection of the resulting infall curves is shown in Figure 2.

Figure 2 shows the normalized shell radius $r/a$ as a function of normalized time $\theta = t/\tau_f$, for six values of the dispersal time scale parameter $\alpha$. For each value of $\alpha$ the radius starts at $r = a$, and decreases to zero at the infall time $\tau_{in}$, or equivalently at the normalized infall time $\theta_{in} = \tau_{in}/\tau_f$. As $\alpha$ decreases, $\theta_{in}$ increases as expected above. Conversely as $\alpha$ approaches the limit $\alpha \to \infty$ (no dispersal), the infall curve exactly matches that of H62, and $\theta_{in} = 1$ as expected. For the range of $\alpha$ which yields $0.3 \leq \epsilon \leq 0.5$, the infall time exceeds the free fall time by 20-40 %, i.e. $1.2 \leq \theta_{in} \leq 1.4$.

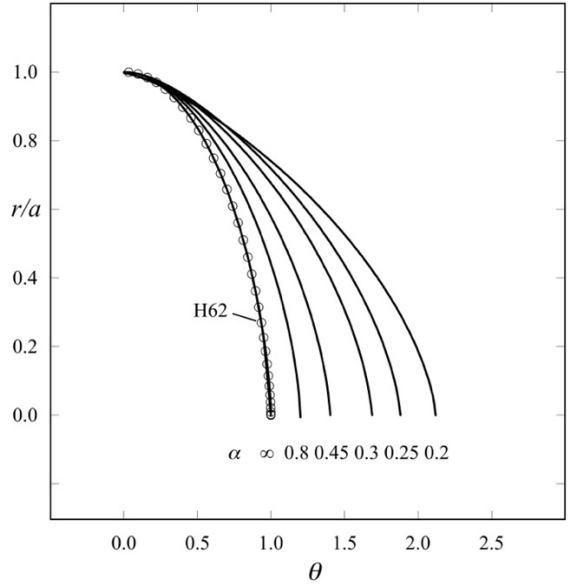

**Figure 2.** Infalling shell radii as a function of time, in six cores with dispersal time scales $\alpha = \infty$ to 0.2 (solid lines from left to right). The radii $r$ are normalized by their initial value $a$, and $\theta$ is the time since the start of infall, normalized by the core free fall time $\tau_f$. The dispersal parameter $\alpha$ equals the dispersal time scale $\tau_d$ normalized by $\tau_f$. Each curve is a solution to the shell EOM, equation (13). The limiting case of $\alpha = \infty$ (no dispersal) coincides with the analytic solution of Hunter (1962), marked H62 (open circles).

## 2.4. Infall Time and Dispersal Parameter

The infall time exceeds the free fall time by the factor $\theta_{in} = \tau_{in}/\tau_f$, which depends only on the dispersal parameter $\alpha$. To predict the evolution of the mass fractions in Section 2.2, equation (13) was solved for a wide range of $\alpha$, and the resulting values of $\theta_{in}$ and $\alpha$ are plotted in Figure 3.

Figure 3 shows a relationship between $\theta_{in}$ and $\alpha$, which is well-fit by the function

$$\theta_{in} = 1 + c_1 \alpha^{-c_2} \qquad (14)$$

with $c_1 = 0.145 \pm 0.001$, $c_2 = 1.300 \pm 0.005$, and correlation coefficient $R > 0.999$ according to the least-squares Levenberg-Marquardt algorithm (Levenberg 1944, Marquardt 1963).

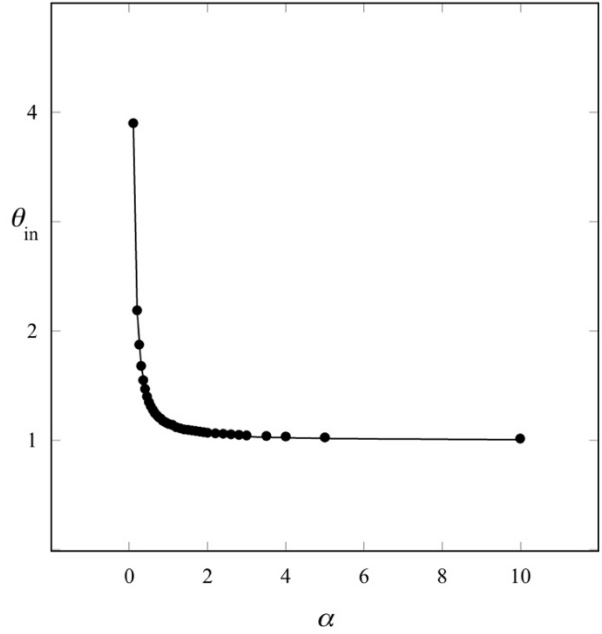

**Figure 3.** Infall times as a function of dispersal parameter, based on solutions to equation (13). The solid line is the best-fit function in equation (14).

## 2.5. Final Mass Fractions

The final mass fractions $\mu_f$ of the protostar, the envelope, and the outflow are evaluated from equation (14) at the time $\theta = \theta_{in}(\alpha)$ when the outermost collapsing shell reaches the origin.

The final mass fraction of the protostar $\mu_{ps,f}$ is equal to the star formation efficiency *SFE*, as discussed in Section 1. Figure 4 shows $\epsilon(\alpha)$ based on substituting the values of $\theta_{in}(\alpha)$ in equation (14) into equation (5). The relation $\epsilon(\alpha)$ is well-fit by a function of the form

$$\epsilon = \frac{1-\exp(-\theta_{in}/\alpha)}{\theta_{in}/\alpha} \qquad (15)$$

as in equation (5), where

$$\theta_{in}/\alpha = \frac{1+c_3 \alpha^{-c_4}}{\alpha}, \qquad (16)$$

with $c_3 = 0.047 \pm 0.001$ and $c_4 = 1.51 \pm 0.03$, based on the Levenberg-Marquardt algorithm (Levenberg 1944, Marquardt 1963) as used in Figure 3. The coefficients in equation (16) are slightly different from those in equation (14), due to the nonlinear dependence of $\theta_{in}$ on $\alpha$.

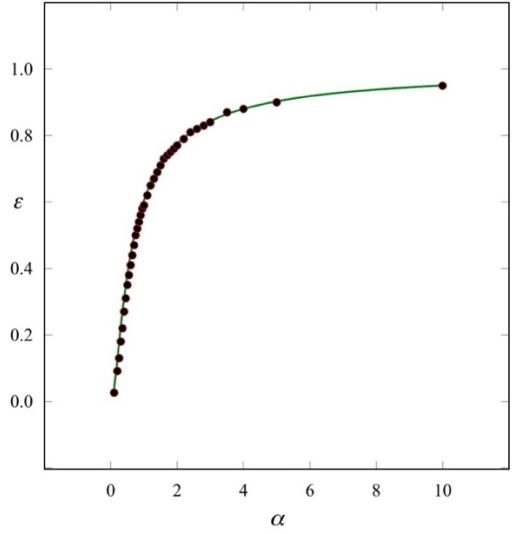

**Figure 4.** Star formation efficiency $\epsilon$ as a function of dispersal parameter $\alpha$, based on solutions to equation (13). The solid line is the best-fit function in equation (15).

The final mass fraction of the envelope, $\mu_{env,f}$, equals zero for all values of $\alpha$, as expected from equation (7) when $\theta = \theta_{in}$. The final mass fraction of the outflow $\mu_{out,f}$, is given in equation (9) as $\mu_{out,f} = 1 - \epsilon$. The final mass fractions for the protostar, envelope, and outflow are shown as functions of $\alpha$ in Figure 5, based on the functions in equations (14)-(16). Figure 5 shows that the standard range of SFE, $\epsilon = 0.30 - 0.50$, is predicted by the range $\alpha = 0.44 - 0.76$.

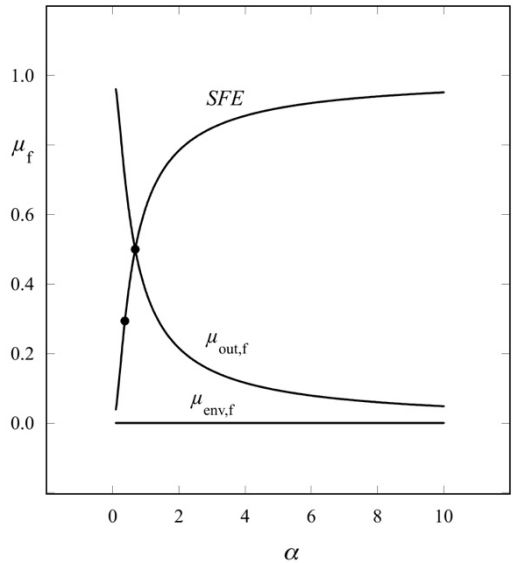

**Figure 5.** Final mass fractions $\mu_f$ with respect to the initial core mass, as functions of the dispersal parameter $\alpha = \tau_d/\tau_f$. The fractions are the final mass of the protostar $SFE = \epsilon = \mu_{ps,f}$, of the infalling envelope $\mu_{env,f}$, and of the outflow $\mu_{out,f}$. These values occur at the time $\tau_{in}$ when the outermost initial shell has accreted onto the protostar, based on equations (14) - (16). *Filled circles* mark the range of $\alpha = 0.44 - 0.76$ corresponding to standard $SFE$ values $\epsilon = 0.30 - 0.50$.

## 2.6. Relative Rates of Dispersal and Accretion

A key property of the IO model is $\omega_f \equiv \overline{(dm/dt)}_{out}/\overline{(dm/dt)}_{ps}$, the ratio of the mean rate of envelope mass dispersal into the outflow to the mean rate of envelope mass accretion onto the protostar. Here each mean is the average over the accretion period $0 \leq \theta \leq \theta_{in}$. This ratio sets the star formation efficiency of the core-protostar system. With final masses defined above in Section 2.5 it is possible to evaluate $\omega_f$ for any $\alpha$ or for any $\epsilon$.

The ratio $\omega_f$ can be evaluated using equation (9), yielding $\omega_f = (1 - \epsilon)/\epsilon$. Over the range of $SFE$ $\epsilon = 0.3 - 0.5$, $\omega_f$ ranges from 1.0 to 2.3. A simple approximation to $\omega_f$ is the time scale ratio $\alpha^{-1}$, which ranges from 1.3 to 2.3 over the same range of $\epsilon$. Thus with either expression for $\omega_f$ the dispersal rate must exceed the accretion rate by a typical factor between 1 and 2 in order to match the expected range of $SFE$.

These estimates of $\omega_f$ includes both the wind component and the entrained component of the outflow gas. The ratio of wind rate to accretion rate is usually estimated to be $0.1 - 0.2$ (Watson et al. 2016; OC17) while the majority of the outflow gas is entrained gas. Observations of protostars in mid-infrared fine-structure lines indicate that $\sim 0.9$ of the mass seen in molecular outflows is matter entrained from the wind surroundings (Watson et al. 2016).

## 3. Evolution of Core Component Masses

This section shows the evolution of the mass fractions of the protostar, envelope, and outflow as functions of time, based on the expressions in Section 2. Section 3.1 shows mass fractions of the protostar, the envelope, and the outflow for six values of the dispersal parameter $\alpha$, in Figure 6 and 7. The results illustrate the importance of the relative infall and dispersal time scales in setting the final mass of the protostar and the durations of the protostellar evolutionary stages. Section 3.2 explains the adopted initial values of core mass and free fall time. It applies these values to predict component masses and evolutionary times for representative values of $\alpha$ and $\epsilon$, in Section 3.3, in Figure 8.

### 3.1. Mass Fraction Evolution

Figures 6 and 7 show mass fraction evolution for a wide range of $\alpha$, based on equations (4), (7), (8), and (10). In each figure a vertical line marks $\theta_0$, the boundary between evolutionary stages 0 and I, and a second line marks $\theta_I$, the boundary between stages I and II. Here the term "stage" is used since it is defined by physical properties, rather than "class" which is defined by observational properties (Dunham et al. 2015).

The normalized "duration" of stage 0 is $\Delta\theta_0 = \theta_0$ and the normalized duration of stage I is $\Delta\theta_I = \theta_I - \theta_0$. Stage 0 is considered the "embedded" phase and stage I is considered the "disk-dominated" phase (Kristensen & Dunham 2018, hereafter KD18). The adopted definitions of $\theta_0$ and $\theta_I$ are chosen for simplicity and for similarity to previous definitions. Here $\theta_0$ is defined as the normalized time when the protostar and envelope masses are equal, marking the end of the main accretion phase (André et al. 2000, Enoch et al. 2008, Offner & Arce 2014, OC17). Here $\theta_I$ is the time $\theta_{in}$ when the outermost shell has reached the origin. At this time the envelope is depleted, so $\theta_I$ approximates the value adopted by MH13, when the envelope mass equals 0.1 of the initial core mass.

The normalized class 0 time $\theta_0$ is closely approximated by $\theta_0 \approx 1/2$. The deviation of $\theta_0$ from 0.50 is less than 0.01, with weak dependence on $\alpha$, based on equations (4), (7), and (14) over the *SFE* range $\epsilon = 0.3 - 0.5$. This property may be understood from equations (4) and (7) and from the infall curves in Figure 2. When $\theta_0 \approx 0.5$, all the infall curves in Figure 2 have nearly the same value as the limiting H62 curve, for which $\theta_{in} = 1$. Then the stage 0 definition $\mu_{ps} = \mu_{env}$ indicates $\exp^{(\theta_0/\alpha)} = 1 + [(1 - \theta_0)/\alpha]$ according to equations (4) and (7). This equation is solved in the limit of small $\theta_0/\alpha$ when $\theta_0 \to 1/2$. The normalized class I time is then $\theta_I = \theta_{in}$ where $\theta_{in}$ is given in equation (14), and the normalized class I duration is $\Delta\theta_I = \theta_{in} - 0.5$. Its typical values are $\Delta\theta_I = 0.7 - 0.9$ over $\epsilon = 0.3 - 0.5$.

In Figures 6 and 7, the mass fractions for different values of $\alpha$ indicate significantly different star-forming outcomes. In Figure 6 when $\alpha \ll 1$, star formation is slow and inefficient, while outflow dispersal is rapid and the outflow mass fraction is high. For example when $\alpha = 0.2$, $\theta_{in} = 3.9$ indicates slow collapse, *SFE* = 0.09 indicates a protostar mass close to that of a brown dwarf for a 1 $M_\odot$ initial core, and $\mu_{out} \approx 0.9$ indicates a strong outflow. In contrast, Figure 7 shows that when $\alpha \gg 1$, star formation is fast and efficient, while outflow dispersal is slow and the outflow mass is negligible. Thus for $\alpha = 5$, $\theta_{in} = 1.02$ indicates fast collapse, $SFE = 0.9$ indicates highly efficient star formation resembling SIS collapse, and $\mu_{out} \approx 0.1$ indicates a weak outflow.

In Figure 6 the protostar accretion histories (*red curves*) show the curvature characteristic of tapered accretion when outflows are important. Similar curvature is seen in numerical studies (MM12, MH13, OC17). In Figure 7, as outflows become less important, the accretion histories become more linear. They approach the limiting case of pure SIS infall, where the accretion rate is constant in time (S77). For $\alpha > 0.45$, the discontinuity in the slope of each evolution curve at $\theta = \theta_{in}$ reflects the idealized discontinuity between the density of the initial core and its truncating medium.

In Figures 6 and 7, the range of $\alpha$ most consistent with standard estimates of $SFE$ is intermediate between the extreme cases, or $\alpha = 0.4 - 0.8$ as noted in Section 2.5

A simple physical derivation of $\alpha$ can be made for a spherically symmetric SIS outflow. The dispersal time scale $\tau_d$ can be expressed as the core-crossing time of the initial gas moving outward at its escape speed. For a SIS truncated at $a$ with no outflow, $\tau_f = \pi a/(4\sigma)$ where $\sigma$ is the thermal velocity dispersion. For spherical outflow from a SIS with no infall, $\tau_d = a/(2\sigma)$. This particular value of $\alpha$ is denoted $\alpha_1 = \tau_d/\tau_f = 2/\pi = 0.64$, independent of $a$ and $\sigma$. This value is used as a "representative" $\alpha$ for $0.3 \leq \epsilon \leq 0.5$ in the evolution curves in Figures 8-9.

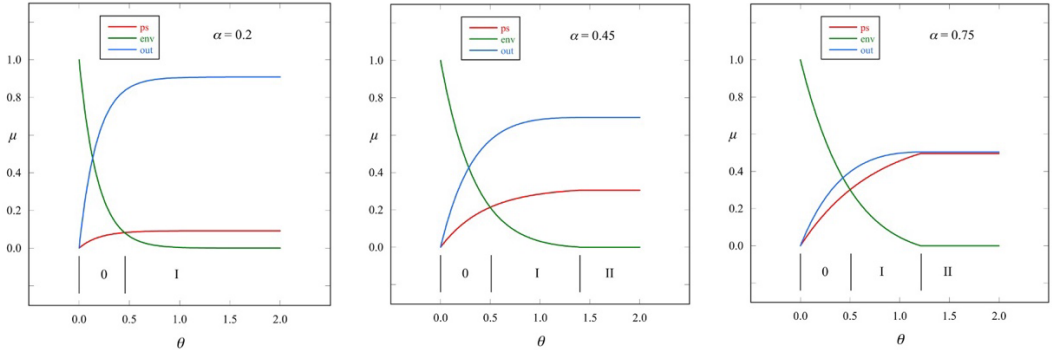

**Figure 6.** Evolution of the core mass fractions $\mu$ in the protostar (*red*), in the infalling envelope (*green*), and in the outflow (*blue*), for fast-dispersing systems with dispersal parameter $\alpha = 0.20, 0.45,$ and $0.75$. Here $\theta = t/\tau_f$ is the ratio of accretion age $t$ to core free fall time $\tau_f$, $\alpha$ is the dispersal parameter $\alpha = \tau_d/\tau_f$, and $\tau_d$ is the dispersal time scale, describing envelope mass loss to the outflow. A vertical line indicates the boundary between evolutionary stages 0 and I, when the protostar and envelope masses are equal, and between stages I and II, when the envelope mass is zero.

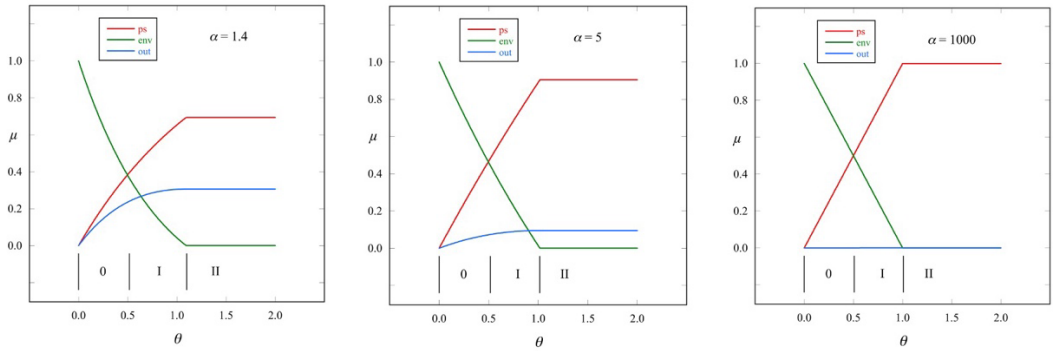

**Figure 7.** Evolution of the core mass fractions $\mu$ in the protostar (*red*), in the infalling envelope (*green*), and in the outflow (*blue*), for slow-dispersing systems with dispersal parameter $\alpha = 1.4, 5.0,$ and $1000$. Here $\theta = t/\tau_f$ is the ratio of accretion age $t$ to core free fall time $\tau_f$, $\alpha$ is the dispersal parameter $\alpha = \tau_d/\tau_f$, and $\tau_d$ is the dispersal time scale describing envelope mass loss to the outflow. As in Figure 6, a vertical line indicates the boundary between evolutionary stages 0 and I, where the protostar and envelope masses are equal, and between stages I and II, where the envelope mass is zero.

### 3.2. Initial Core Mass and Free Fall Time

The foregoing mass fractions and evolutionary times are normalized, for application to a range of initial masses and time scales. This section selects particular values of core initial mass and free-fall time, based on observations and simulations. These choices allow comparison of the model predictions with observed protostar masses, with estimated durations of evolutionary stages, and with observed outflow opening angles.

The initial core mass is $M(0) \approx 1\, M_\odot$, consistent with *Herschel* studies of starless (prestellar) cores in nearby star-forming regions, including Perseus (Sadavoy et al. 2014, Mercimek et al. 2017), Orion B (Könyves et al. 2020), and Serpens (Fiorellino et al. 2021). For specificity the value $M(0) = 1.05\, M_\odot$ is adopted, to match the initial mass of seven star-forming cores with outflows in detailed MHD simulations (MH13).

The initial core free-fall time is probably in the range 80-100 kyr according to several estimates. Observational estimates include $\tau_f = 100 \pm 22$ kyr, the mean ± standard deviation of free fall times in 69 *Herschel* starless dense cores in Perseus (Mercimek et al. 2017) and $\tau_f \approx 90$ kyr from outflow properties (B96). The MHD simulations of MH13 have mass 1.05 $M_\odot$, temperature 10 K, and $\tau_f = 82$ kyr. The SIS model and an initial transonic velocity dispersion 0.27 km s$^{-1}$ based on NH$_3$ line observations of starless dense cores (Chen et al. 2019) indicate $\tau_f = 91$ kyr and truncation radius 6500 au for the same initial mass and temperature as in MH13. For consistency with the MH13 value, the fiducial time scale $\tau_f = 82$ kyr is adopted here. However we caution that this time scale could be as long as ~100 kyr.

### 3.3. Mass Evolution for Representative Parameters

The mass and time scales adopted above in Section 3.2 are used with the representative time scale ratio $\alpha_1 = 2/\pi$ described in Section 3.1. They show the evolution of core component masses in both normalized and dimensional form in Figure 8. This choice of $\alpha_1 = 0.64$ is intermediate between the values of $\alpha$ assumed in Figure 5, $\alpha = 0.45$ (center) and $\alpha = 0.75$ (right).

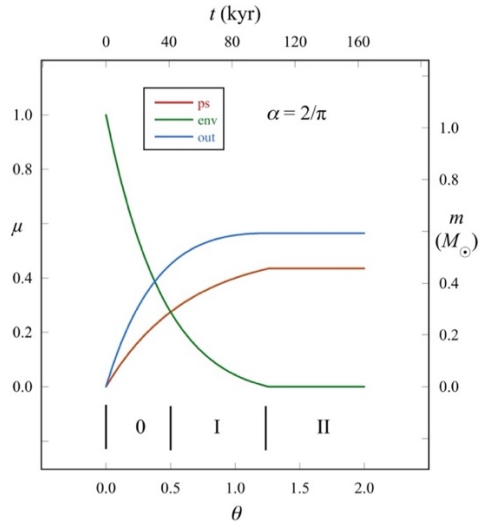

**Figure 8.** Evolution of core mass fractions $\mu$ in the protostar (*red*), in the infalling envelope (*green*), and in the outflow (*blue*), for a protostellar system with representative dispersal parameter $\alpha_1 = 2/\pi$. The resulting $\mu_{ps,f} = \epsilon = 0.44$ lies within the usually estimated range $\epsilon = 0.3\text{-}0.5$. As in Figures 6 and 7, $\theta = t/\tau_f$ is the ratio of accretion age $t$ to core free fall time $\tau_f$. The accretion age (*top x-axis*) is based on free-fall time $\tau_f = 82$ kyr, and the mass scale (*right y-axis*) is based on initial core mass $M(0) = 1.05\ M_\odot$. A vertical line indicates the adopted boundary between evolutionary stages 0 and I, when the protostar and envelope masses are equal, and between stages I and II, when the envelope mass is zero.

In Figure 8, the *SFE* is $\epsilon = 0.44$, within the standard range of *SFE* values. The protostar mass at the end of stage 0 is 0.28 $M_\odot$, the final protostar mass is 0.46 $M_\odot$, and the final outflow mass is 0.59 $M_\odot$. The evolutionary stage boundaries are at $t_0 = 41$ kyr and $t_I = 103$ kyr, and the stage durations are $\Delta t_0 = 41$ kyr and $\Delta t_I = 62$ kyr.

## 4. Outflow Opening Angles

This section estimates final opening angles $\phi_{oaf}$ of observed outflow cavities. It derives the cavity volume fraction for power-law cavity shapes, and relates the volume fraction to estimates of the *SFE*. It uses these results to predict the opening angle evolution for an expanding paraboloidal cavity.

### 4.1. Outflow Mass to Volume Ratio

As noted in Section 2.2, the *SFE* can be written in terms of the final outflow mass fraction $\mu_{out,f}$ as

$$\epsilon = 1 - \mu_{out,f}. \qquad (17)$$

Equation (17) assumes that the envelope loses mass only to the protostar and to the outflow, and that it does not gain mass once the infall and outflow begin. If so, the final outflow mass fraction must be $\mu_{out,f} = 0.5 - 0.7$ for consistency with the standard estimates of $\epsilon = 0.3 - 0.5$.

Direct estimation of the mass outflow fraction $\mu_{out,f}$ from line observations has uncertainties in molecular abundance and excitation, line optical depth, and confusion between envelope and core gas at low velocities (Dunham et al. 2014, hereafter D14). Estimates from infrared continuum observations are also limited by uncertain inclination, radiative transfer, and cavity evolution (H21).

It is therefore useful to consider the final outflow cavity volume fraction $v_{\text{out,f}} \equiv V_{\text{out,f}}/V(0)$, defined as the ratio of the final outflow cavity volume within the initial core boundary $V_{\text{out,f}}$, to the initial core volume $V(0)$. The ratio $v_{\text{out,f}}$ can be estimated more directly from observations than can the outflow mass fraction $\mu_{\text{out,f}}$. Then equation (17) can be rewritten as

$$\epsilon = 1 - \eta_f v_{\text{out,f}} \quad , \tag{18}$$

where the "outflow mass to volume ratio" $\eta_f \equiv \mu_{\text{out,f}}/v_{\text{out,f}}$ is the normalized ratio of the final outflow mass to the final outflow cavity volume. Here $v_{\text{out,f}}$ can be estimated from observations, and equation (18) can determine values of $\eta_f$ needed for consistency with the expected range of $\epsilon$.

If $\eta_f = 1$, the final outflow mass equals the product of the mean initial core density and the final cavity volume within the initial sphere boundary. This property is expected for an outflow whose mass grows at the same rate as its core-bounded cavity volume. This equality of rates can occur if envelope gas is entrained into the outflow along the expanding cavity walls (Z16, Z19). On the other hand, $\eta_f > 1$ implies that the outflow cavity volume grows more slowly than the outflow mass. This difference in rates may occur when envelope gas is entrained only near the disk and when lateral expansion of the cavity walls is limited, possibly by the core magnetic field (MM12, MH13).

### 4.2. Outflow Survey Observations

Several imaging studies allow estimates of the final outflow cavity volume fraction $v_{\text{out,f}}$. It is usually assumed that the final opening angle is closest to the largest opening angle in an observed sample, i.e. $\phi_{\text{oaf}} \approx \phi_{\text{oam}}$, and similarly for the corresponding volume fractions $v_{\text{out,f}} \approx v_{\text{out,max}}$ (H21, D23). These are approximations due to statistical uncertainty and because once the envelope is fully depleted, the cavity walls within the core boundary are not detectable. However these approximations are somewhat justified because at late times the opening angle increases relatively slowly, as shown in observations (D23) and in Figure 9.

Many protostellar outflows have been mapped in $^{12}$CO (1-0) and $^{12}$CO (2-1) line emission at millimeter wavelengths (Langer et al. 1996, A06, D23). The largest CO study (D23) estimated opening angles with high or medium confidence from 37 outflows, primarily from Class 0 protostars in Perseus (Tobin et al. 2016, Stephens et al. 2018, 2019). Scattered-light continuum images have also been analyzed in the mid-infrared (MIR) from 31 Spitzer IRAC observations at 3.6 μm to 8.0 μm wavelength of regions within a few hundred pc of the Sun (Seale & Looney 2008, V14). In the largest near-infrared (NIR) outflow study with the Hubble Space Telescope, 29 scattered-light outflows were imaged at 1.6 μm wavelength from protostars in Orion (H21).

### 4.3. Estimates of Maximum Outflow Angle

The maximum angle $\phi_{oam}$ in a sample of outflow observations is important to quantify angle expansion as a probe of protostar evolution, as noted above. This section estimates and discusses $\phi_{oam}$ based on the V14, D23, and H21 surveys.

To quantify the angle distribution for each survey, a hyperbolic tangent fit was made to a plot of outflow opening angle $\phi_{oa}$ as a function of bolometric temperature $T_{bol}$. The fit function is $\phi_{oa} = \phi_{oam} \tanh[(T_{bol} - T_{bol,shift})/T_{bol,0}]$, as in D23. Here $T_{bol,shift}$ sets the horizontal position of the function and $T_{bol,0}$ sets the value of $T_{bol} - T_{bol,shift}$ where the function slope changes from steep to shallow. The asymptotically flat part of each fit gives a statistical estimate of $\phi_{oam}$. These are $\phi_{oam}(V14) = 108 \pm 10$ deg, $\phi_{oam}(D23) = 80 \pm 9$ deg, and $\phi_{oam}(H21) = 49 \pm 5$ deg. The first two of these fits match those in D23 Table 5, but here the one-sigma fit uncertainties in $\phi_{oam}$ are also reported. The fits have correlation coefficients $R(V14) = 0.64$ for 31 points, $R(D22) = 0.55$ for 37 points, and $R(H21) = 0.48$ for 29 points.

### 4.3.1. MIR and CO Comparison

The $\phi_{oam}$ angle difference of ~29 deg between the V14 MIR sample and the D23 CO sample is corroborated by the median angle difference $\Delta\phi_{oa} = 25$ deg among the four sources common to the samples (D23). Well-resolved images of the HH 46/47 outflow system also show a similar angle difference when the MIR image is superposed on a high-resolution ALMA CO image (A13, Z16 Figure 16a).

These systematic MIR-CO angle differences may arise because widest part of the outflow has the lowest velocities, where the $^{12}$CO line cannot trace structure because of its significant optical depth (A13, V14, D14). This explanation is supported by ALMA observations of low-velocity emission in HH 46/47 in the optically thin C$^{17}$O line. There the wider-angle optically thin C$^{17}$O emission matches the MIR image more closely than the narrower-angle optically thick CO emission (Z16 Figure 16b). Also, the MIR enhanced-resolution images have mean resolution 0.8 arcsec (V14). These MIR images may be able to resolve some cavity edges more clearly than the CO images, which have mean resolution 2.3 arcsec (D23).

In some cases, the D23 CO fit angle does not coincide with the IO model angle where the outflow cavity crosses the core boundary, as described in the Appendix. However, the associated angle uncertainty is less than the typical uncertainty in estimating the D23 angle, and is significantly less than the typical MIR-CO angle difference.

An estimate of $\phi_{oam}$ was also made for a combination of the MIR and CO samples, assuming that the CO angles systematically underestimate the MIR angles. The angle offset $\phi_{oam}(V14) - \phi_{oam}(D23)$ was added to each D23 angle before combining with the V14 angles. The resulting combined fit gave $\phi_{oam} = 108$ deg, as expected for consistency with a systematic angle offset. However the correlation coefficient was less than for either fit alone, indicating significantly different distributions within each sample.

### 4.3.2. CO and NIR Comparison

The D23 CO fits have maximum outflow angles which exceed the H21 NIR maximum angles at the same values of $T_{bol}$, again by $\Delta\phi_{oa} \approx 30$ deg. These CO angles may exceed the NIR angles because some of the CO emission comes from molecular gas entrained outside the NIR cavity (H21, Seale & Looney 2007). Some of the H21 outflows are deeply embedded, so that their extended emission may suffer significant NIR extinction. It is also possible that the H21 observations are sensitive to a population of smaller cavities which are not detected by the other surveys. It will be important for future observations to observe a common sample of protostars at the same wavelengths, rather than the present samples which have few sources in common.

### 4.3.3. Angle Summary

In summary, the best statistical estimate of the above angle upper limits $\phi_{oam}$ is probably $\phi_{oam}(V14) = 108 \pm 10$ deg based on the above tanh fits. This angle of 108 deg is also a strict upper limit to 34 of the 37 CO angles in D23, and to all 29 of the NIR angles in the H21 sample.

Such maximum angles $\phi_{oam}$ near 110 deg exceed the prediction $\phi_{oam} \approx 90$ deg due to entrainment of the outflow gas near the disk and channeling of the outflow by the core magnetic field (MM12, MH13). In contrast the wide-angle outflow models of A13, Z16, and Z19 entrain envelope gas along their expanding cavity walls, and have no such limitation on their expansion due to the core magnetic field.

## 4.4. Outflow Cavity Shapes

Each of the V14, D23, and H21 surveys has a significant fraction of images with one to four curved "arms" of concave shape, with each arm tracing a cavity wall. This fraction increases with improving resolution, and models of the best-resolved images are well-fit by parabolic shapes.

Visual inspection of Figures 3, 4, and 5 in the D23 CO survey indicates 16 sources where clear assignments of arm shape can be made. Of these, at least six sources have two clear concave arms. In the V14 MIR survey, visual inspection of Figure 4 indicates that 16 sources have two or more concave arms, confirming the V14 shape description as "typically parabolic." In the H21 NIR survey, power-law shape fits were made of the form $z \propto x^n$ where $x$ is the half-width perpendicular to the cavity symmetry axis, and $z$ is the height along the axis. Here $n$ is the degree of the power law, with $n = 1$ for cones and $n = 2$ for parabolas. For 29 fits the power-law degree $n$ ranges from 1.0 to 6.7, with mean 1.9 and median 1.5.

In summary, more than one-third of the D23 CO images have curved features suggestive of parabolic shapes, which are typical of cavity images in the finer-resolution MIR and NIR surveys. Indeed parabolic CO shapes are prominent in finer-resolution ALMA CO images of HH 46/47 (A13, Z16, Z19).

## 4.5. Outflow Cavity Volume Fractions

This section gives an expression for the outflow cavity volume fraction $v_{out} \equiv V_{out}/V(0)$ as a function of angle $\phi$ and power-law degree $n \geq 1$. Here $v_{out}$ is the volume fraction whose final value $v_{out,f}$ is defined in Section 4.1. This expression is used to evaluate the outflow mass to volume ratio and the corresponding $SFE$ due to different power-law cavity shapes.

The volume fraction $v_{out}$ can be expressed as the volume fraction of a bipolar solid of revolution centered in a bounding sphere. The cavity volume $V_{out}$ occupies a fraction $v_{out} \equiv V_{out}/V(0)$ of the initial core volume $V(0)$. A radial line from the origin to the intersection of the cavity with the sphere lies at a polar angle $\phi$ from the symmetry axis. Here $\phi = \phi_{oa}/2$ where $\phi_{oa}$ is the full opening angle shown in Figure 1.

The cavity volume fraction can then be written as

$$v_{out} = 1 - a_n c_\phi - b_n c_\phi^3 \qquad (19)$$

where $c_\phi \equiv \cos\phi$ and the coefficients are $a_n \equiv 3/(n+2)$ and $b_n \equiv (n-1)/(n+2)$. Thus $a_1 = 1$ and $b_1 = 0$ for cones, and $a_2 = 3/4$ and $b_2 = 1/4$ for paraboloids.

Equation (19) is derived by calculating the volume of revolution for each of the two unipolar cavities whose shape has degree $n \geq 1$. Each of these cavities is truncated by the plane through its intersection with the spherical boundary of the initial core. The volume of each truncated cavity is added to the volume of its joining spherical cap. Then the total bipolar cavity volume is divided by the volume of the initial sphere to obtain $v_{out}$. Equation (19) is more general than volume cavity expressions for a cone or paraboloid because it also applies to non-integer values of $n$, and to values $n > 2$.

Equation (19) indicates that the cavity volume fraction increases with shape degree, as shown in H21 Figure 3 for cones and paraboloids. The volume fraction of a shape with $n > 1$ is equal to that of a cone ($n = 1$) only at the extreme angles $\phi = 0$ and $\phi = \pi/2$. At intermediate angles the volume fraction difference $\Delta v \equiv v_n - v_1$ has a local maximum $\Delta v_m \equiv (1 - a_n - b_n/3)/\sqrt{3}$ when $\phi \equiv \phi_0 = \cos^{-1}(1/\sqrt{3})$. The magnitude of $\Delta v_m$ increases with increasing $n$, while $\phi_0$ is the same for all $n > 1$, including paraboloids ($n = 2$). The opening angle corresponding to $\phi_0$ is $\phi_{oa0} \equiv 2\cos^{-1}(1/\sqrt{3}) = 109.5$ deg.

The angle $\phi_{oa0}$ matches within uncertainty to the maximum observed opening angle $\phi_{oam} = 108 \pm 10$ deg estimated in Section 4.3 above. This coincidence is convenient because $\cos(\phi_{oa0}/2) = 1/\sqrt{3}$ provides a simple expression for evaluation of equations (19) and (22). At this angle equation (19) indicates that the cavity volume fraction $v_{out}$ increases significantly with power-law degree $n$. Thus $v_{out} = 0.42$ for a cone ($n = 1$), $v_{out} = 0.52$ for a paraboloid ($n = 2$), and $v_{out} = 0.62$ for a quartic shape ($n = 4$).

### 4.6. Estimates of *SFE* for Wide-angle Cavities

The opening angle observations of H21 and D23 indicate values of *SFE* which exceed the standard values $\epsilon = 0.3 - 0.5$ when analyzed with frequently made assumptions, as noted in Section 1. The *SFE* is $\epsilon \approx 0.7$ when $\phi_{oam} \approx 90$ deg as predicted by MH13, when the cavity shape is conical, and when the normalized outflow mass-to-volume ratio is $\eta_f = 1$. Equivalently, to match $\epsilon = 0.3 - 0.5$ with a conical cavity shape requires the normalized outflow mass to volume ratio to have the range $1.7 \leq \eta_f \leq 2.4$, a significant increase over the expected value $\eta_f \approx 1$.

This section shows that estimates of *SFE* have much closer agreement with $\epsilon = 0.3 - 0.5$ if the maximum opening angle $\phi_{oam}$ has a "wide-angle" value near 110 deg as in Section 4.3, and if the cavity shape is similar to that of a paraboloid, or to a shape of greater power-law degree as in Section 4.4. Assuming $\phi_{oam} = \phi_{oa0}$ and $\eta_f = 1$, equations (18) and (19) indicate $\epsilon = 0.58$ for a conical cavity shape and $\epsilon = 0.48$ for a paraboloidal shape. Thus increasing the maximum angle beyond ~90 deg brings $\epsilon$ closer to the range 0.3 - 0.5 for a conical shape. However both a wider angle and a shape resembling a paraboloid are needed to bring $\epsilon$ into the range 0.3 - 0.5. To match the entire range $\epsilon = 0.3 - 0.5$, the mass to volume ratio must lie in the range $1.2 \leq \eta_f \leq 1.7$ for a cone, or in the range $1.0 \leq \eta_f \leq 1.4$ for a paraboloid. Thus paraboloidal cavities match expected values of both $\epsilon$ and $\eta_f$ more closely than do conical cavities.

At the maximum opening angle $\phi_{oa0}$ the outflow cavity can have a paraboloidal shape over the entire range $\epsilon = 0.3 - 0.5$, with relatively small departures of $\eta_f$ from unity. Substitution of $\phi = \phi_{oa0}/2$ and $n = 2$ into equations (18) and (19) yields the relation $\eta_f = (1 - \epsilon)/[1 - 5/(6\sqrt{3})]$. Then as $\epsilon$ decreases from 0.5 to 0.4 to 0.3, $\eta_f$ must slightly exceed unity, from 1.0 to 1.2 to 1.4. Similarly, $\eta_f$ can be held to within the range $1.0 - 1.1$ if final outflow cavities have paraboloidal shape ($n = 2$) for $\epsilon = 0.5$ and quartic shape ($n = 4$) for $\epsilon = 0.4$ and $\epsilon = 0.3$.

In summary, if outflow cavity angles increase to final values $\gtrsim 110$ deg with cavity shapes of degree $\gtrsim 2$, their associated values of *SFE* can range over the expected range $\epsilon = 0.3 - 0.5$, with nearly equal ratios of outflow mass and cavity volume, normalized to initial core values. Such wide maximum angles and nearly parabolic shapes match expected values of *SFE* more closely than do narrow maximum angles and conical shapes.

The foregoing relations between $\phi_{oaf}$ and *SFE* apply to maximum angles $\phi_{oaf}$ but not necessarily to smaller values $\phi_{oa} < \phi_{oaf}$, since the *SFE* is defined only for final values of the outflow mass and volume fractions. A model of the increase in outflow angle with time over the range $0 \leq \phi_{oa} \leq \phi_{oaf}$ is presented in Section 4.7.

### 4.7. Outflow Angle Evolution

To estimate how outflow angles evolve, a model of a radially expanding paraboloidal shell (Li & Shu 1996, Lee et al. 2000, A13, Z16, Z19) is used to obtain the evolution of the volume fraction of a bipolar paraboloidal shell within its initial sphere.

For a paraboloidal shell whose symmetry axis lies along the z-axis in the $x - z$ plane, the height $z$ of a point on the shell is related to its cylindrical radius $x$ by $(z/R_0) = (x/R_0)^2$, where $R_0$ is the scale length of the paraboloid. For this scale length, the polar angle $\phi$ from the z-axis to a point $(x, 0, z)$ on the paraboloid is $\phi = \tan^{-1} \sqrt{R_0/z}$, so that $x = z = R_0$ when $\phi = \pi/4$.

When the origin of the paraboloid coincides with the center of a sphere of radius $R_S$ as in Figure 1, the paraboloid equation $(z/R_0) = (x/R_0)^2$ and the sphere equation in the plane of the sky $x^2 + z^2 = R_S^2$ can be combined to yield two equivalent expressions. The first gives $R_0/R_S$ in terms of the polar angle $\phi = \phi_{oa}/2$,

$$\frac{R_0}{R_S} = \frac{(\sin \phi)^2}{\cos \phi} . \qquad (20)$$

The second expression gives $\phi$ in terms of $R_0/R_S$,

$$\phi = 2 \tan^{-1} \left\{ \frac{2}{[1+(2R_S/R_0)^2]^{1/2}-1} \right\}^{1/2} . \qquad (21)$$

Equations (20) and (21) are used with the expansion of the paraboloidal scale length $R_0$ to predict the evolution of the outflow opening angle, in terms of time and of the outflow final angle.

ALMA observations of CO emission from the protostellar system HH46/47 indicate that the outflow structure can be well-fit by multiple paraboloidal shells which are launched episodically (A13, Z16, Z19). The shells are interpreted as expanding linearly, so that each has a scale length $R_0$ increasing linearly with time $t$. As they expand, new shells catch up with old shells to form a single expanding cavity wall. Here the cavity wall is modeled as a single expanding shell.

For such linear expansion the paraboloid scale length can be written $R_0 = R_{0f}\, t/\tau_{in}$ where $R_{0f}$ is the scale length at the final time of the outflow shell expansion. This final time is assumed to equal the time $\tau_{in}$ when the accretion ends since the accretion powers the expansion. Combining with equation (20) yields the scale length in terms of time and final angle, as $R_0 = R_S(t/\tau_{in})\,(\sin\phi_f)^2/\cos\phi_f$.

The final paraboloid has final angle $\phi_f = \phi_{oam}/2 = \cos^{-1}\sqrt{1/3}$, assuming $\phi_{oam} = \phi_{oa0}$ as discussed in Section 4.5. Then substitution of $R_0 = R_S(t/\tau_{in})(2/\sqrt{3})$ into equation (21) yields the full opening angle $\phi_{oa}$ in terms of time $t$,

$$\phi_{oa} = 2\tan^{-1}\left\{\frac{2}{[1+3(\tau_{in}/t)^2]^{1/2}-1}\right\}^{1/2}. \qquad (22)$$

This expression is verified by evaluating equation (22) at the initial time $t = 0$, yielding $\phi_{oa}(0) = 0$, and at the final time $t = \tau_{in}$, yielding $\phi_{oam} = 2\tan^{-1}\sqrt{2}$ rad $= 2\cos^{-1}\sqrt{1/3}$ rad $= 109.5$ deg, as expected.

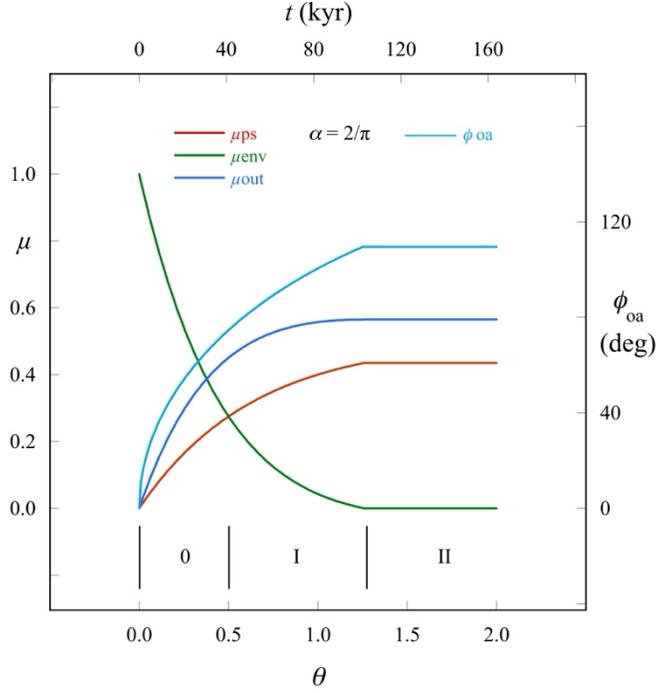

**Figure 9.** Evolution of outflow opening angle $\phi_{oa}$ *(light blue)* and of core mass fractions μ in the protostar (*red*), in the infalling envelope (*green*), and in the outflow (*blue*), for a protostellar system with representative dispersal parameter $\alpha = 2/\pi$. As in Figures 6-8, $\theta = t/\tau_f$ is the ratio of accretion age $t$ to initial core free fall time $\tau_f$. The accretion age (*top x-axis*) is based on free-fall time $\tau_f = 82$ kyr. Vertical lines indicate the adopted boundary $\theta_0$ between evolutionary stages 0 and I, when the protostar and envelope masses are equal, and $\theta_I$ between stages I and II, when the envelope mass is zero. During stage II each quantity is assumed to keep its final value at $\theta_I$.

The evolution of the opening angle $\phi_{oa}$ calculated from equation (22) for a paraboloidal cavity is shown in Figure 9. This figure also shows the mass fractions of the protostar, the infalling envelope, and the outflow for the representative dispersal parameter $\alpha_1 = 2/\pi$, as in Figure 8. For this value of $\alpha$ the *SFE* is $\epsilon = 0.44$, within the standard range of *SFE* values. The mass-to-volume ratio is then $\eta_f = 1.1$, also close to the standard value $\eta_f = 1.0$.

## 5. Comparison with Observations

This section compares the foregoing IO model predictions with observed values of protostar final mass fractions and final masses, evolutionary stage durations, outflow opening angles, and the evolutionary stage assignments of these opening angles.

*Masses and SFE.* The range of dispersal parameter $0.4 \lesssim \alpha \lesssim 0.8$ which predicts *SFE* $0.3 \lesssim \epsilon \lesssim 0.5$ also predicts final protostar masses $0.3\ M_\odot \lesssim m_f \lesssim 0.5\ M_\odot$ for the adopted initial core mass $M(0) = 1.05\ M_\odot$. These masses lie within 50% of the mean mass of the initial mass function (IMF), $\bar{m}_{\mathrm{IMF}} = 0.36\ M_\odot$ (Weidner & Kroupa 2006). The closest match occurs when $\alpha = 0.50$, $m_f = \bar{m}_{\mathrm{IMF}}$ and $\epsilon = 0.35$. Thus these model properties and parameters are consistent with characteristic stellar masses and characteristic values of the *SFE*.

*Stage Durations.* The adopted core free-fall time $\tau_f = 82$ kyr sets the time scale for model estimates of the class 0 duration $\Delta\tau_0$ and of the class I duration $\Delta\tau_I$. For the representative dispersal parameter $\alpha_1 = 2/\pi = 0.64$, the predicted durations are $\Delta\tau_0 = 41$ kyr and $\Delta\tau_I = 62$ kyr, as shown in Figure 7. These durations are similar to the durations $\Delta\tau_0 = 47 \pm 4$ kyr and $\Delta\tau_I = 88 \pm 7$ kyr derived from a "half-life" analysis of young stellar objects in nearby Gould Belt clouds (Evans et al. 2009, Dunham et al. 2015, Kristensen & Dunham 2018, hereafter KD18).

The predicted durations for $\alpha_1 = 2/\pi$, $\Delta\tau_0 = 41$ kyr and $\Delta\tau_I = 62$ kyr, are also similar to the typical durations in the above MHD simulations. The mean durations over five simulations with the same initial mass and free-fall time as in Section 3.2 are $\Delta\tau_0 = 29$ kyr and $\Delta\tau_I = 84$ kyr (MH13 Table 2). However the apparent agreement between model class durations and half-life durations has uncertainty noted in Section 3.2, due to a spread of estimates of $\tau_f$ extending up to $\sim 100$ kyr.

In contrast, durations estimated from relative population counts in each class are longer than half-life durations by a factor ~3, i.e. $\Delta\tau_0 \approx 150$ kyr and $\Delta\tau_I \approx 300$ kyr (Dunham et al. 2015), as discussed in Section 6.

*Outflow Angles and Stages.* Figure 9 shows that the predicted time behavior of $\phi_{oa}$ appears consistent with observed outflow angles and evolutionary stages. The selected value of $\phi_{oam} = 110$ deg is consistent with the statistical estimate of the maximum MIR angle observed by V14. It is also an upper limit to 34 of the 37 CO angles observed by D23, and to all of the 29 NIR angles observed by H21, as noted in Section 4.3. In a model of expanding paraboloidal outflow cavities, $\phi_{oa}$ increases with time most rapidly during the stage 0 phase, and $\phi_{oa}$ approaches its final value more slowly during the stage I phase. This behavior is also seen in Figure 9 of D23, when the bolometric temperature $T_{bol}$ is considered a proxy for early evolutionary times. The predicted opening angle reaches the median observed value $\phi_{oa} = 54$ deg (D23) at the normalized time $\theta = 0.25$, or about half-way through the stage 0 duration. This median angle is observed when $T_{bol}$ is close to the midpoint of its stage 0 values, consistent with $\theta \approx 0.25$.

## 6. Model Summary
### 6.1. Model Properties

Recent observations have raised the question whether protostellar outflows can remove enough dense core mass to clear cores and to set the masses of low-mass stars. This paper investigates the conditions under which outflows can match observations and remove enough gas mass. A model of an evolving protostellar core finds the necessary ratio of outflow mass to protostar mass, and the ratio of outflow mass to outflow cavity volume, to match typical values of stellar mass, star formation efficiency, and outflow opening angle. The model matches statistical estimates of the durations of the embedded (stage 0) and disk-dominated (stage I) phases of young stellar object evolution.

The method used is an "infall-outflow" (IO) model of a star-forming dense core. The initial core has the density structure of a truncated SIS. It forms a protostar by collapsing pressure-free and by losing mass to the outflow. Each infalling mass shell loses a fraction $\exp(-t/\tau_d)$ of its initial mass to the outflow, where $t$ is the time since the start of infall and $\tau_d$ is a time scale for dispersal of envelope gas into the outflow, as in B96. This time scale $\tau_d$ is shorter than the initial free fall time $\tau_f$ by a typical factor $\alpha^{-1} \equiv \tau_f/\tau_d \approx 2$. The initial core has transsonic microturbulence, and its rotation and magnetic field strength are small compared to equipartition values. Nonetheless they are sufficient to form a protostellar disk and to launch a wide-angle bipolar outflow driven by a hydromagnetic wind.

The protostar mass accretion rate is based on the pressure-free collapse of a truncated SIS. The rate is $M(0)\tau_{in}^{-1}\exp(-t/\tau_d)$ where $M(0)$ is the initial core mass and $\tau_{in}$ is the "infall time" for the outermost mass shell to reach the origin. This rate is tapered because an increasing fraction $[1-\exp(-t/\tau_d)]$ of the initial envelope mass is entrained into the outflow, leaving a decreasing fraction of envelope mass available to accrete onto the protostar. The protostar mass is obtained by integrating the accretion rate over time, and the final protostar mass is evaluated at $t = \tau_{in}$. The *SFE* equals the ratio of the final protostar mass to the initial core mass. Similar derivations are made for the mass fractions of the infalling envelope and of the outflow in Section 2.

The infall time $\tau_{in}$ exceeds the free fall time $\tau_f$, since mass loss to the outflow continuously decreases the inward gravitational force exerted on each infalling shell. For this mass-losing system the shell equation of motion (EOM) has no analytic solution, in contrast to the constant-mass case solved in H62. Numerical solutions to the EOM are shown in Figure 2. They match the H62 solution in the limit $\alpha \to \infty$, and they show that the infall time $\tau_{in}$ increases as $\alpha$ decreases. For the *SFE* range $\epsilon = 0.3 - 0.5$ the infall time exceeds the free fall time by a factor $\tau_{in}/\tau_f = 1.2 - 1.4$.

The dependence of $\tau_{in}$ on $\alpha$ is shown in Figure 3, and the dependence of *SFE* on $\alpha$ is shown in Figure 4, based on solutions to the shell EOM. The dependence of the final mass fractions of the protostar, the envelope, and the outflow on $\alpha$ are shown in Figure 5. These indicate that the range of $\alpha = 0.4 - 0.8$ yields the range of *SFE* $\epsilon = 0.3 - 0.5$. The corresponding ratio of final outflow mass to protostar mass is then 1.0 - 2.3.

The time evolution of the core component mass fractions is presented in Figures 6-8 for a wide range of $\alpha$. These figures show low-*SFE* outcomes for fast dispersal in Figure 6 and high-*SFE* outcomes for slow dispersal in Figure 7. They also show the relation of mass fractions to evolutionary stages. The boundary between stages 0 and I is the time when the protostar and envelope masses are equal, as in OC17, at time $\tau_f/2$. The boundary between stage I and II is the time $\tau_{in}$ when accretion ends, approximating the time adopted by MH13.

Mass and time scales are assigned to the dimensionless evolution model, following $M(0) = 1.05\ M_\odot$ and $\tau_f = 82$ kyr for consistency with observations and simulations. The resulting mass evolution curves are shown in Figure 8 for the representative dispersal parameter $\alpha_1 = 2/\pi$. The evolutionary durations are 41 kyr for the embedded stage 0 period and 70 kyr for the disk-dominated stage I period.

The IO model predictions are compared with observations of stellar masses, star formation efficiency, and class durations. For the range of $\alpha$ which matches $\epsilon = 0.3 - 0.5$, the final protostar masses lie within 50% of the mean mass of the IMF. The rate of outflow mass increase exceeds the rate of protostar mass increase by a factor 1.0 - 2.3. The predicted class durations are similar to the estimated half-life durations of Gould Belt YSOs (KD18).

Model outflow angles increase up to a maximum which approximates the greatest observed angles $\phi_{oam} \approx 110$ deg in surveys of CO line maps and scattered-light images by D23, V14, and H21. The outflow shell is a radially expanding paraboloid which entrains envelope gas, as in the Z19 model of ALMA observations of HH 46/47. For the range of $\alpha$ matching $\epsilon = 0.3 - 0.5$, the mean rate of outflow mass gain exceeds the mean rate of cavity volume gain by a factor $\eta_f = 1.0 - 1.4$. This factor is significantly less than in earlier estimates, which were based on smaller values of $\phi_{oam}$ and on a conical cavity shape. The range of $\eta_f$ can be further reduced to $\eta_f = 1.0 - 1.1$ if the final cavity shape is paraboloidal when $\epsilon = 0.5$ and quartic when $\epsilon = 0.3 - 0.4$. The predicted outflow angle increases rapidly during the stage 0 phase, where it matches observed angles near 50 deg. It increases more slowly during the stage I phase, reaching its maximum value near 110 deg after ~100 kyr.

With the foregoing outflow rates and cavity shapes the IO model matches observed protostar masses, *SFE*s, class durations, and outflow angles, with no need for additional mechanisms of gas dispersal.

### 6.2. Model Parameters

Table 1 summarizes the basic parameters of the model. The initial core mass $M(0)$, free fall time $\tau_f$, velocity dispersion $\sigma$, and dispersal parameter $\alpha$ are used with the equations in sections 2 and 3 to predict the mass evolution of the protostar, envelope, and outflow, and the durations of the evolutionary stages 0 and I. These show consistency with the expected range of the *SFE* and with observational estimates of stage durations. The outflow final opening angle $\phi_{oaf}$ and the cavity shape degree $n$ are used with the equations in section 4 to predict the evolution of opening angles. These predictions approximate observed opening angles, and they indicate that cavity volumes trace outflow masses within a typical factor 1.2 over the expected range of the *SFE*.

**Table 1**

**Model Parameters**

| (1) Name | (2) Symbol | (3) Value | (4) Section of first use |
|---|---|---|---|
| initial core mass | $M(0)$ | 1.05 $M_\odot$ | 3.2 |
| velocity dispersion | $\sigma$ | 0.27 km s$^{-1}$ | 3.2 |
| free fall time | $\tau_\mathrm{f}$ | 82 kyr | 3.2 |
| dispersal parameter | $\alpha$ | 0.64 | 3.1 |
| final opening angle | $\phi_\mathrm{oaf}$ | 110 deg | 4.5 |
| outflow shape degree | $n$ | 2 | 4.6 |

Note - Listed parameter values of $M(0), \sigma, \tau_\mathrm{f}$, and $\phi_\mathrm{oaf}$ are fixed, while values of $\alpha$ and $n$ are each representative of a range of values discussed in the text.

## 7. Discussion
### 7.1. Implications

The main implication of this work is that protostellar outflows can clear their host cores and set protostar masses. If so, there is no need to invoke additional mechanisms of core gas dispersal to account for observed estimates of *SFE*. This result can be ascribed to two changes from some earlier estimates.

First, well-resolved outflow surveys in the MIR, the NIR, and in CO lines indicate that the widest observed outflows have opening angles at least $\phi_\mathrm{oa} \approx 110$ deg (Section 4.3), significantly wider than the ~90 deg angles predicted in some simulations (MH12, MH13). Second, well-resolved outflow images often have concave structure (Section 4.4). They are better described by paraboloidal power-law shapes of degree $n \gtrsim 2$ than by conical shapes of degree $n = 1$, as in early outflow studies with coarser resolution. Such paraboloidal cavities enclose greater volume than cones of the same opening angle.

With these two changes, wide-angle final outflow cavities of nearly paraboloidal shape can clear enough volume to match *SFE* values in the range $0.3 \leq \epsilon \leq 0.5$, provided the final outflow mass-to-volume ratio is limited to $1.0 \leq \eta_f \leq 1.4$. In contrast, conical outflows with $\phi_{oam} \approx 90$ deg would require $1.7 \leq \eta_f \leq 2.4$ to match the same range of *SFE*. Such large ratios of outflow mass to volume appear inconsistent with simple models of outflow expansion, as noted previously (H21, D23).

A second implication is that the evolution of protostar masses and outflow angles to reach their final observed values can be understood quantitatively in a simple model of competition between gravitational infall and outflow expansion. However this IO model is restricted to idealized initial conditions, and it applies only to formation of a single low-mass star from a non-accreting core, as discussed below.

## 7.2. Limitations and Uncertainties
### 7.2.1. Initial Density Structure

The main limitation of the IO model is the simplicity of its assumptions, motivated by the goal of an analytic approach. These assumptions include the initial concentration of the SIS, which appears physically artificial (Whitworth & Summers 1985). The error in the SIS density structure is greatest near the origin, where the model also neglects the detailed structure of the disk where the outflow is launched. Thus the model envelope density structure is not applicable for radii less than ~100 au, which includes the inner few percent of the initial envelope mass.

The IO model neglects the effects of magnetic field strength and rotation on the core size scale due to their relatively small energies. These initial relative energies are assumed to be similar to those assumed by MM12 and MH13. These authors assume that the magnetic energy is typically 0.06 of the gravitational energy, and that the rotational energy is typically 0.01 of the gravitational energy (MH13 Table 1). This combination is sufficient to launch significant hydromagnetic wind and outflow motions from the disk scale once the protostar forms. Nonetheless these low energy ratios indicate a small departure from the initial envelope structure on the core scale, according to models where the SIS is modified with low levels of magnetization or rotation.

In the singular isothermal toroid (SIT), the magnetized equilibrium analog of the SIS, the initial mass to flux ratio μ ~7 corresponds to a small magnetization parameter $H_0 < 0.06$ (Li & Shu 1996 Table 1). This value of $H_0$ implies that the SIT density at a given radius departs by at most a few percent from the SIS density at the same radius. This departure factor is bounded by the "density function" $R$ in Li & Shu (1996) Figure 1a, for the case $n = 0.25$.

Observational estimates of mass to flux ratio in low-mass dense cores are typically in the range μ = 1-3, indicating stronger fields than the initial values assumed here (Crutcher 2012; Myers & Basu 2021). However these estimates are mostly based on regions which have embedded protostars, so their mass to flux ratios can not be considered as initial conditions for collapse. The relation of the initial μ prior to collapse to the present-day μ derived from observations remains to be studied in more detail. The evolution of weak initial fields to stronger values has been seen in some simulations, as turbulent fragmentation amplifies the field (Li et al 2015).

An equilibrium solution for a slowly rotating SIS with velocity dispersion $\sigma$ and angular rotation rate $\Omega$ has density at a given radius $r$ which exceeds that of a nonrotating SIS by the factor $\exp(\Phi)$, where $\Phi$ is a power series in $\xi^2$ whose leading term is $\Phi = \xi^2/4$, and where $\xi = r\Omega/\sigma$ (Terebey, Shu & Cassen 1984). For the rotation rate $\Omega = 10^{-14}$ s$^{-1}$ and mass $M = 1.05 \, M_\odot$ assumed by MH13, $\xi = 0.103$ and the density factor $\exp(\Phi)$ is less than 1.01.

Thus the assumed low initial ratios of magnetic and rotational energy to gravity indicate that the initial densities of the modified SIS differ from those of the true SIS by at most a few percent on the core scale. This result supports the assumption of SIS initial densities in the IO envelope model.

The frequent association of a dense core with surrounding filamentary gas (André et al. 2014) implies that some collapsing cores have significant filamentary departure from spherical symmetry. If so, it may be more difficult for outflows to remove their infalling dense gas than in the spherically symmetric case considered here. The importance of this effect should be quantified by numerical studies, which are outside the scope of this paper.

### 7.2.2. Turbulence, Core Environment, and Protostar-Disk System

The effects of turbulent motions on the collapse can lead to significant differences in multiplicity and outflow shape between the MHD simulation results of Offner et al. (2016) and OC17, who include turbulent motions, and MM13, who neglect them. Thus the results of the IO model apply to cores with modest turbulence, and only to some features of turbulent cores, as noted below in Section 7.4.

The isolation of the collapsing system from its environment in the IO model is unrealistic because it does not allow for accretion from the surrounding medium, which can alter the infall dynamics and can lengthen the infall duration (Kaminski et al. 2014). However the core mass gain due to formation accretion is relatively small, about 10% during the protostar collapse time, according to models of prestellar core formation from converging large-scale flows in a turbulent, magnetized medium (Chen & Ostriker 2014, hereafter CO14). The median core mass formed is ~0.47 $M_\odot$ in ~1 Myr (CO14). A similar growth time of ~1 Myr was found in a model of dense core formation by gravitational instability of a filament (Heigl et al. 2016). Thus in these simulations the typical core formation mass accretion rate is ~$5 \times 10^{-7}$ $M_\odot$ yr$^{-1}$. This rate may be too low to double the core mass within a core lifetime. Such a rate was suggested to help account for the similarity of core and star turnover masses in Orion (Takemura et al. 2023). In contrast, the typical protostar mass accretion rate in the IO model is greater by a factor ~10, $8\sigma^3/(\pi G) \approx 5 \times 10^{-6}$ $M_\odot \text{yr}^{-1}$ according to section 2.1 when $T = 10$ K.

The assumed core isolation also neglects gas dispersal due to external stellar feedback in the form of winds, ionization and neighboring outflows (e.g. Tanaka et al. 2017). These dispersal mechanisms can be important in regions of massive star formation, but the main result of this paper suggests that such external feedback is not necessary to account for most core and protostar properties.

In this work the protostar and disk are considered a single entity, as in MM00, but this assumption neglects the different timing of accretion onto the protostar from the envelope and from the disk. In MM13 the disk forms before the protostar and the two have significantly different accretion histories. Thus IO model protostar mass estimates less than ~0.1 $M_\odot$ are more uncertain than for greater masses.

### 7.2.3. Evolutionary Stage Durations

The model durations of the stage 0 and stage I phases agree well with the estimates for more than 100 YSOs in Gould Belt clouds, according to the half-life analysis of KD18. However the agreement is significantly worse if instead the counting method is used to analyze the same population data (KD18, D15). The counting method assumes that the relative class populations in each stage are in steady state, while the half-life method assumes a constant birthrate and a constant probability of decay from one stage to the next. The half-life durations are then shorter than the counting durations because the half-life of each stage needs to be short enough for a significant fraction of the class II protostars to have already passed sequentially through the earlier stages (KD18).

Model class durations are also uncertain because the assumed half-life duration of the class II stage, 2 Myr, could be as long as ~ 3 Myr according to some estimates (D15). This would not change the ratio of duration estimates between half-life and counting methods, because each is scaled to the class II duration. Nonetheless, the KD18 half-life estimates 47 and 88 kyr assuming that the class II duration equals 2 Myr are corroborated by their similarity to the mean duration estimates in the MH13 simulations, 29 and 84 kyr.

### 7.2.4. Resolution Limitations

The IO model identifies two conditions for good model agreement with observations, and these conditions have partial agreement with the MHD simulations of MM12 and MH13. However these simulations do not resolve the outflow jet, and thus can not reveal the full details of how the outflow is composed of jet, wind, and entrained gas. Also, the IO model of exponential mass loss from spherical shells is a highly simplified description of the actual envelope mass loss to the outflow. It remains to be determined whether the conditions identified here continue to hold, when compared with more detailed simulations and observations made with finer resolution.

### 7.3. Comparison to Similar Studies

Analytic outflow models with hydromagnetic wind properties have been combined with infall models for a magnetized singular isothermal toroid (Li & Shu 1996), where the wind is confined to a cone of fixed angular width (MM00). The IO and MM00 models each consider a collapsing spherical core with outflow driven by a magnetized wind, with the protostar and disk as a single object. They differ because the IO model calculates the infall time based on the reduction of internal density due to the outflow, while MM00 assume that the wind does not slow accretion. The IO mass accretion rate is tapered with time while in MM00 it is constant in time. The IO model predicts the time evolution of the protostar, envelope, and outflow masses and the durations of evolutionary classes, while MM00 predict only final masses. The IO model predicts that the outflow angle widens with time to a maximum ~110 deg within one free-fall time, while MM00 predict a fixed angle of ~130 deg for typical parameters. The IO model matches $SFE = 0.3 - 0.5$ in accord with recent observations while MM00 predict $SFE = 0.25$-$0.75$.

Among analytic models, the IO model is most similar to that in Myers (2008, hereafter M08), where a spherically symmetric envelope embedded in an extended medium loses mass at an exponentially declining rate, while it undergoes pressure-free collapse. However the model of M08 does not match detailed properties of protostars, envelopes and outflows which have become known in recent years. These include observed outflow opening angles limited to $\lesssim 110$ deg (V14, D23); estimates of *SFE* in the range $\epsilon \approx 0.3 - 0.5$ as noted above; and estimates of the half-life durations of the evolutionary classes $\tau_0 \approx 50$ kyr and $\tau_I \approx 90$ kyr (KD18).

A model based on a rotating, collapsing SIS (Terebey et al. 1984) was developed to match line and continuum observations of the protostellar core B335 (Evans et al. 2023). The best-fit velocity dispersion, collapse age, and protostar mass are similar to those modelled here, while the envelope outer radius is greater by a factor ~3. An outflow cavity is included to match observations although its evolution is not modeled. This study is the most detailed available, approximating high-resolution observations in numerous molecular lines and at continuum wavelengths.

Among numerical simulations, the IO model closely resembles MH13 by design in initial masses, and time scales. The models are also similar in *SFE* and evolutionary stages, as noted earlier. They differ because the IO model considers the disk and protostar as a single entity while they are separate in MH13, with different accretion rates. They also differ because MH13 require significant nonradial motions to entrain envelope gas into the base of the outflow. In contrast, the expanding wide-angle paraboloidal cavity in the IO model entrains envelope gas along the cavity wall and therefore has a smaller component of nonradial motions.

The IO model also resembles the simulations of OC17, in initial core mass, size, and temperature. The OC17 model has significantly greater initial turbulent motions than the IO model. OC17 have initial magnetic fields but no rotation, in contrast to MM12 and MH13, who assume rotation and magnetic fields but no turbulence. OC17 employ a subgrid model of the outflow based on MM00. They predict the evolution of protostar mass, jet mass and entrained mass as functions of time, with evolution curves in their Figure 8 depending on the initial mass-to-flux ratio of the core. The IO curves in Figures 5 and 6 resemble these OC17 curves in their shapes and in their values of *SFE*. However the OC17 durations are 2-3 times longer than the IO durations, resembling counting method durations (D15) more closely than half-life durations (KD18).

## 8. Conclusion

Recent observations of protostellar outflow structure have raised the question whether a typical outflow can remove enough mass from a dense core to match observed estimates of core-star efficiency. This question is addressed with a model of time-dependent infall and outflow ("IO model") from a dense core with fixed initial mass $\approx 1\ M_\odot$. The model is summarized in section 6.

The main results of this paper are:

1. Protostellar dense cores evolve toward their final outflow angles and their final protostar and outflow masses, according to a simple model of infall and outflow. The initial core resembles a truncated SIS with small fractions of magnetic, rotational, and turbulent energy. Spherical shells of envelope gas fall radially inward while bipolar outflow shells of paraboloidal shape expand radially outward. The envelope loses mass to the protostar with free fall time scale $\tau_f$ and to the outflow with dispersal time scale $\tau_d$. The dispersal parameter $\alpha = \tau_d/\tau_f$ sets the *SFE* $\epsilon$ and the ratio of infall and free fall times.

2. The IO model predicts *SFEs* within the usually estimated range $\epsilon = 0.3 - 0.5$ for $\alpha = 0.4 - 0.8$. The mean rate of outflow mass gain is then ~twice the mean rate of protostar mass gain. The model approximates the mean mass of the IMF for initial core masses near $1\ M_\odot$. It predicts evolutionary stage durations to be $0.5\tau_f \approx 40$ kyr for the embedded stage 0 and $0.9\tau_f \approx 70$ kyr for the disk-dominated stage I, approximating half-life duration estimates of Gould Belt YSOs.

3. The protostellar outflow is described as a linearly expanding bipolar shell of paraboloidal shape, based on high-resolution mid-infrared and CO observations. Its opening angle on the core scale approaches ~110 deg at the end of the protostellar accretion phase. Such wide-angle paraboloids and cavities of similar shape clear enough dense core volume and mass to match $\epsilon = 0.3 - 0.5$, in contrast to estimates based on narrower angles and conical shapes.

4. Outflow opening angles are predicted to increase rapidly through ~50 deg during stage 0 and then more slowly as they approach ~110 deg during stage I. These angles and evolutionary times approximate observed CO outflow angles and their associated evolutionary stages.

5. Despite the model simplifications, the matches of predicted and observed masses, *SFE*s, evolutionary stages, and outflow angles indicate no need for external mechanisms of envelope dispersal in setting the formation properties of single, low-mass stars.

**Acknowledgements**

The ideas in this paper were developed from data obtained with the Submillimeter Array (SMA) through the "MASSES" Legacy Program. The authors are grateful to the SMA Director Ray Blundell and the SMA staff for their extensive support of this project. We also wish to recognize and acknowledge the very significant cultural role and reverence that the summit of Maunakea has always had within the indigenous Hawaiian community. We are most fortunate to have had the opportunity to use observations from this mountain. The authors thank Tyler Bourke, Mark Krumholz, and Stella Offner for helpful discussions. We thank the referees for helpful suggestions which improved the paper.

**Appendix.** Comparing two different methods for measuring outflow opening angles

In this paper, the opening angle of a parabolic outflow is measured as the full opening angle of a cone that is centered on the driving source and extends through the point where the parabolic outflow intersects the 6500 au core boundary. Given that we make extensive comparisons to the outflow opening angles reported by the MASSES project (D23), in this appendix we compare opening angles measured as defined here with those measured using the method of D23 which is based on fitting a Gaussian to the distribution of angles to all pixels within the outflow (see D23 for details).

To perform this comparison we generated 15 parabolic outflows of different widths using the equation $y = Ax^2$, with the parameter $A$ used to vary the width. The values of $A$ were chosen to generate outflows spanning the full range of observed outflow widths. For each parabolic outflow we calculated $\phi_{oa,p}$, the opening angle as defined in this paper. We then applied the exact same fitting method used by D23 to determine $\phi_{oa,m}$, the opening angle that would have been reported had this outflow been observed by the MASSES project. Table 2 below presents the results of this comparison, and Figure 10 shows three representative examples for outflows with $A = 0.002$ au$^{-1}$, $A = 0.0004$ au$^{-1}$, and $A = 0.0001$ au$^{-1}$.

**Table 2**

**Measured Opening Angles for Parabolic Outflows**

| $A$ (au$^{-1}$) | $\phi_{oa,p}$ (degrees) | $\phi_{oa,m}$ (degrees) | $\phi_{oa,m} - \phi_{oa,p}$ (degrees) |
|---|---|---|---|
| 0.01 | 14.3 | 18.2 | 3.9 |
| 0.005 | 20.2 | 25.7 | 5.5 |
| 0.003 | 26.1 | 24.9 | -1.2 |
| 0.002 | 31.7 | 28.4 | -3.3 |
| 0.0015 | 36.5 | 34.6 | -1.9 |
| 0.0012 | 40.8 | 41.6 | 0.8 |
| 0.001 | 44.6 | 43.7 | -0.9 |
| 0.0008 | 49.7 | 43.7 | -6 |
| 0.0006 | 56.8 | 52.6 | -4.2 |
| 0.0005 | 62.1 | 62.4 | 0.3 |
| 0.0004 | 68.7 | 62.6 | -6.1 |
| 0.0003 | 78.2 | 67.7 | -10.5 |
| 0.0002 | 93.3 | 85.3 | -8 |
| 0.0001 | 121.1 | 111.1 | -10 |
| 0.00005 | 145.6 | 146 | 0.4 |

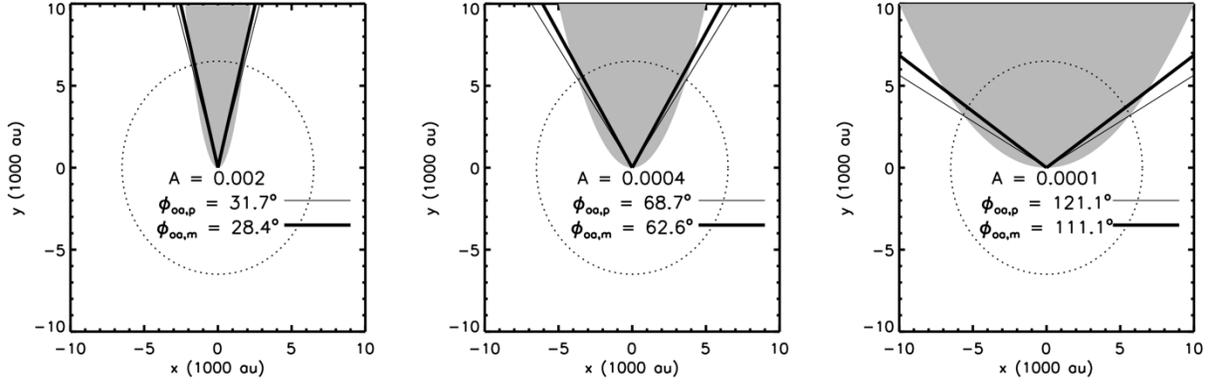

**Figure 10.** Images showing parabolic outflows defined as $y = Ax^2$, with $A = 0.002$ au$^{-1}$ (left), $A = 0.0004$ au$^{-1}$ (center), and $A = 0.0001$ au$^{-1}$ (right). The shaded gray areas show the parabolic outflows, the thin solid lines show cones with opening angles measured as defined in this paper, and the thick solid lines show cones with opening angles measured using the fitting procedure described by D23. The dashed circle in each panel shows the assumed core boundary of 6500 au. Only one lobe of each outflow is presented in order to leave space for the annotations in each panel.

In general, the agreement in outflow opening angles between the two methods is excellent. The mean difference between the two methods, calculated as $\phi_{oa,m} - \phi_{oa,p}$, is -2.7°, and the standard deviation of the difference between the two methods is 4.7°. As the two methods agree both to within one standard deviation, and within the typical uncertainties of 5° – 15° quoted by D23 for their method, we conclude that it is appropriate in this paper to compare to the MASSES opening angles measured by D23.

**References**

Alves, J., Lombardi, M., & Lada, C. 2007, A&A, 462, L17

André, P., Di Francesco, J., Ward-Thompson, D. et al. 2014, in Protostars and Planets VII, eds. H. Beuther, R. Klessen, C. Dullemond, & T. Henning (Tucson: Univ. Arizona Press), 27

Arce, H., & Sargent, A. 2006, ApJ, 646, 1070 (AS06)

Arce, H., Mardones, D., Corder, S. et al. 2013, ApJ 774, 39 (A13)

Arce, H., Shepherd, D., Gueth, F. et al. 2007, in Protostars and Planets V, ed. B. Reipurth, D. Jewitt, & K. Keil (Tucson: Univ. Arizona Press), 245

Bally, J. 2016, ARA&A, 54, 49

Bontemps, S., André, P., Terebey, S. et al. 1996, A&A, 311, 858 (B96)

Chandrasekhar, S. 1939, An Introduction to the Study of Stellar Structure (Chicago: University of Chicago Press), 84

Chen, C.-Y., & Ostriker, E. 2014, ApJ, 785, 69

Chen, H., Myers, P., Ladd, E. et al. 1995, ApJ, 445, 377

Cunningham, A., Krumholz, M., McKee, C. et al. 2018, MNRAS, 476, 771

Curtis, E., Richer, J., Swift, J. et al. 2010, MNRAS, 408, 1516

Dunham, M., Arce, H., Mardones, D. et al. 2014, ApJ, 783, 29 (**D14**)

Dunham, M., Allen, L., Evans, N. et al. 2015, ApJS, 220, 11

Dunham, M., Stephens, I., Bourke, T. et al. 2022, ApJ, submitted (D23)

Enoch, M., Evans, N., Sargent, A. et al. 2008, ApJ, 684, 1240

Evans, N., Dunham, M., Jørgensen, J. et al. 2009, ApJS, 181, 321

Evans, N., Yang, Y., Green, J. et al. 2023, ApJ, 943, 90

Fiorellino, E., Elia, D., André, P. et al. 2021, MNRAS, 500, 4257

Frank, A., Ray, T., Cabrit, S. et al. 2014, in Protostars and Planets VI, ed. H. Beuther, R. Klessen, C. Dullemond, & T. Henning (Tucson: Univ. Arizona Press), 451

Habel, N., Megeath, S., Booker, J. et al. 2021, ApJ, 911, 153 (H21)

Hansen, C., Klein, R., McKee, C. et al. 2012, ApJ, 747, 22

Hosokawa, T., Omukai, K., Yoshida, N. et al. 2011, Sci, 334, 1250

Heigl, S., Burkert, A., & Hacar, A. 2016, MNRAS, 463, 4301

Hsieh, T-H., Lai, S-P., & Belloche, A. 2017, AJ, 153, 173

Hunter, C. 1962, ApJ, 136, 594 (H62)

Kaminski, E., Frank, A., Carroll, J. et al. 2014, ApJ, 790, 70

Könyves, V., André, P., Men'shchikov, A. et al. 2015, A&A, 584, 91

Könyves, V., André, P., Arzoumanian, D. et al. 2020, A&A, 635, 34

Kristensen, L., & Dunham, M. 2018, A&A, 618, 158 (KD18)

Krumholz, M. 2015, Notes on Star Formation, arXiv:1511.03457, 103

Kuiper, R., Turner, N., & Yorke, H. 2016, ApJ, 832, 40

Lada, C. 2006, ApJ, 640, L63

Langer, W., Velusamy, T., & Tie, T. 1996, ApJ, 468, L41

Lee, C.-F., Mundy, L., Reipurth, B. et al. 2000, ApJ, 542, 925

Levenberg, K. 1944, QApMa, 2, 164

Li, P., McKee, C., & Klein, R. 2015, MNRAS, 452, 2500

Li, Z.-Y., & Shu, F. 1996 ApJ, 472, 211

Machida, M. & Matsumoto, T. 2012, MNRAS, 421, 588 (MM12)

Machida, M., & Hosokawa, T. 2013, MNRAS, 431, 1719 (MH13)

Marquardt, D. 1963, SJAM, 11, 431

Matzner, C. & McKee, C. 2000, ApJ, 545, 364 (MM00)

Mercimek, S., Myers, P., Lee, K. et al. 2017, AJ, 153, 214

Motte, F., André, P., Neri, R. 1998, A&A, 336, 150

Myers, P. 2008, ApJ, 687, 340 (M08)

Myers, P., & Basu, S. 2021, ApJ, 917, 35

Nakano, T., Hasegawa, T., & Norman, C. 1995, ApJ, 450, 183

Offner, S., Lee, E., Goodman, A. et al. 2011, ApJ, 743, 91

Offner, S., & Arce, H. 2014, ApJ, 784, 61

Offner, S., Dunham, M., Lee, K. et al. 2016, ApJ, 827, L11

Offner, S., & Chaban, J. 2017, ApJ, 847, 104 (OC17)

Offner, S., Taylor, J., Markey, C. et al. 2022a, MNRAS, 517, 88

Offner, S., Moe, M., Kratter, K. et al. 2022b, arXiv:2203.10066

Polyanin, A., & Zaitsev, V. 2018, Handbook of Ordinary Differential Equations
    (Boca Raton: CRC Press)

Pudritz, R., & Norman, C. 1986, ApJ, 301, 571

Rohde, P., Walch, S., Clarke, S. 2021, MNRAS 500, 35

Schmeja, S., & Klessen, R. 2004, A&A, 419, 505

Seale, J., & Looney, L. 2008, ApJ, 675, 427

Shu, F. 1977, ApJ, 214, 488 (S77)

Snell, R., Loren, R., & Plambeck, R. 1980, ApJ, 239, L17

Spitzer, L. 1978, Physical Processes in the Interstellar Medium (New York: Wiley), 286

Stephens, I., Dunham, M., Myers, P. et al. 2018, ApJS, 237, 22

Stephens, I., Bourke, T., Dunham, M. et al. 2019, ApJS, 245, 21

Takemura, H., Nakamura, F., Arce, H. et al. 2023, ApJS, 264, 35

Tanaka, K., Tan, J., & Zhang, Y. 2017, ApJ, 835, 32

Terebey, S., Shu, F., & Cassen, P. 1984, ApJ, 286, 529.

Tobin, J., Looney, L., Li, Z-Y. et al. 2016, ApJ, 818, 73

Tobin, J., Offner, S., Kratter, K. et al. 2022, ApJ, 925, 39

Velusamy, T., Langer, W., Thompson, T. 2014, ApJ, 783, 6 (V14)

Vorobyov, E. 2010, ApJ, 713, 1059

Watson, D., Calvet, N., Fischer, W. et al. 2016, ApJ, 828, 52

Weidner, C., & Kroupa, P. 2006, MNRAS, 365, 1333

Whitworth, A., & Summers, D. 1985, MNRAS, 214, 1

Zhang, Y., Arce, H., Mardones, D. et al. 2016, ApJ 832, 158 (Z16)

Zhang, Y., Arce, H., Mardones, D. et al. 2019, ApJ, 883, 1 (Z19)